\documentclass[preprint,12pt]{elsarticle}
\usepackage{graphicx}
\usepackage{bm}
\usepackage{color}
\usepackage{amssymb}
\usepackage{amsmath}

\newcommand{\BCAO}{BaCo${_{2}}$(AsO${_{4}}$)${_{2}}$}

\begin{document}


\title{Polarized-neutron investigation of magnetic ordering and spin dynamics in BaCo${_{2}}$(AsO${_{4}}$)${_{2}}$ frustrated honeycomb-lattice magnet}

\author[ill]{L.-P.~Regnault\corref{cor1}} 
\ead{louis-pierre.regnault@orange.fr}
\author[bnp]{C.~Boullier}
\author[in]{J.E.~Lorenzo}

\cortext[cor1]{Corresponding author}

\address[ill]{Institut Laue Langevin, 71 avenue des Martyrs, 38042 Grenoble cedex 9, France}
\address[bnp]{ BNP-Paribas, 20 boulevard des Italiens, 75009 Paris, France}
\address[in]{Institut N\'eel-CNRS/UJF, F-38042 Grenoble, Cedex 9, France}

\date{\today}

\begin{abstract}
The magnetic properties of the cobaltite {\BCAO}, a good realization of the quasi two-dimensional frustrated honeycomb-lattice system with strong planar anisotropy, have been reinvestigated by means of spherical neutron polarimetry with CRYOPAD. From accurate measurements of polarization matrices both on elastic and inelastic contributions as a function of the scattering vector {\bf{Q}}, we have been able to determine the low-temperature magnetic structure of
{\BCAO}  and reveal its puzzling in-plane spin dynamics. Surprisingly, the ground-state structure (described by an incommensurate propagation vector ${\bf{k}}_{1}=(k_{x}, 0, k_{z}$), with $k_{x}=0.270{\pm}0.005$ and $k_{z} \approx -1.31$) appears to be a quasi-collinear structure, and not a simple helix, as previously determined. In addition, our results have revealed the existence of a non-negligible out-of-plane moment component $ \approx 0.25{\mu}_{B}$/Co$^{2+}$, representing about 10\% of the in-plane component, as demonstrated by the presence of finite off-diagonal elements $P_{yz}$ and $P_{zy}$ of the polarization matrix, both on elastic and inelastic magnetic contributions. Despite a clear evidence of the existence of a slightly inelastic contribution of structural origin superimposed to the magnetic excitations at the scattering vectors ${\bf{Q}}=(0.27, 0, 3.1)$ and ${\bf{Q}}=(0.73, 0, 0.8)$ (energy transfer ${\Delta}E \approx 2.3$ meV), no strong inelastic nuclear-magnetic interference terms could be detected so far, meaning that the nuclear and magnetic degrees of freedom have very weak cross-correlations. The strong inelastic $P_{yz}$ and $P_{zy}$ matrix elements can be understood by assuming that the magnetic excitations in {\BCAO} are spin waves associated with trivial anisotropic precessions of the magnetic moments involved in the canted incommensurate structure.

\end{abstract}


\maketitle


\section{Introduction}
\label{intro}

\begin{figure}
\centering
\includegraphics[width=10cm]{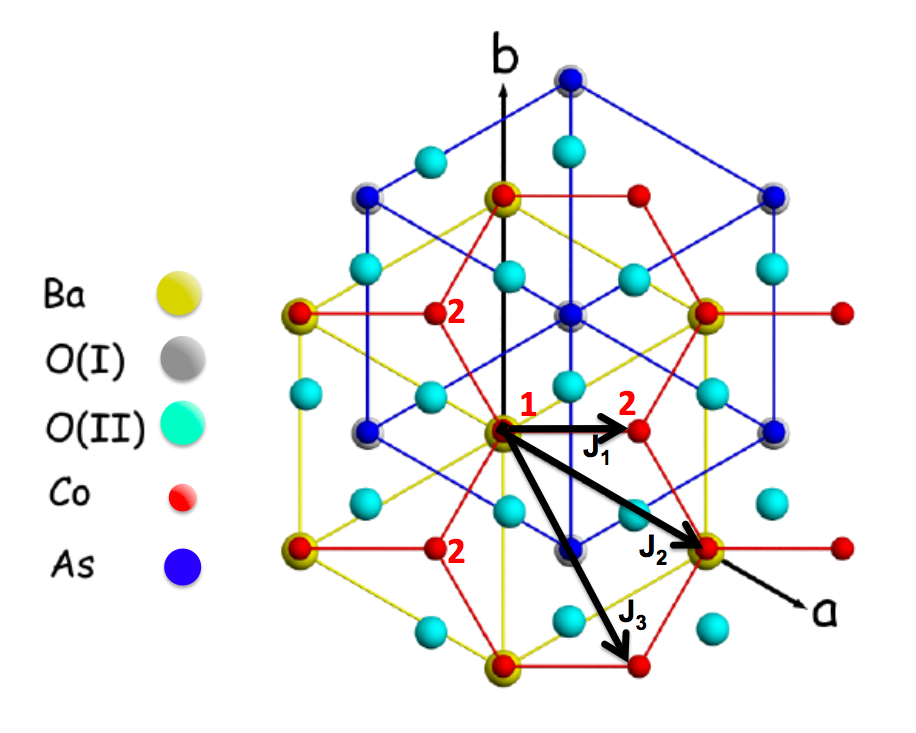}
\caption{In-plane projection of the crystallographic structure of {\BCAO}, showing the honeycomb lattice of Co$^{2+}$ ions and the two hexagonal Bravais sublattices (labeled 1 and 2). $J_{1}$, $J_{2}$ and $J_{3}$ are respectively the exchange-coupling constants between the first, second and third neighbors on the hexagons.}
\label{Figure-StructCristallo}
\end{figure}

\begin{figure}
\centering
\includegraphics[width=8cm]{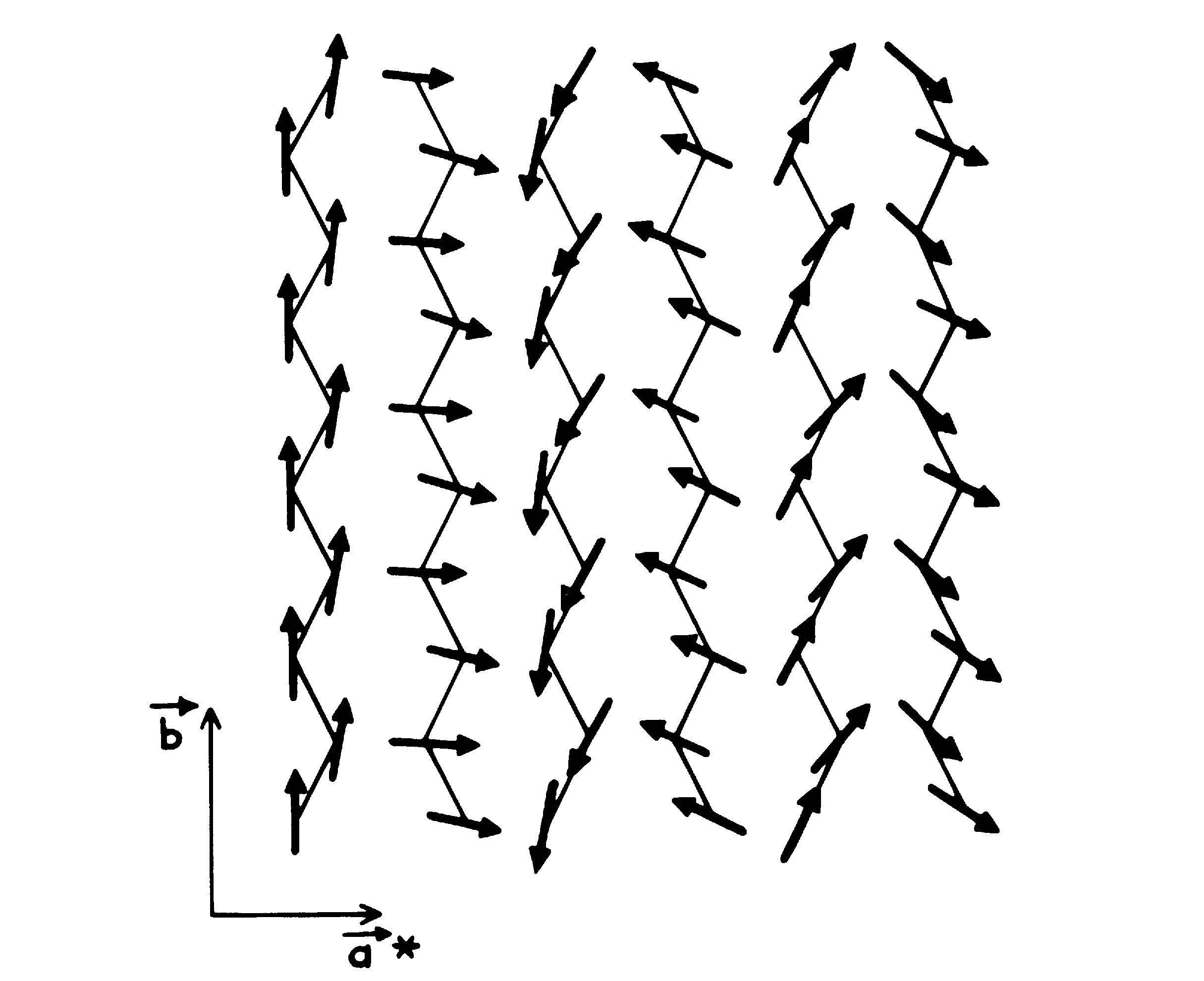}
\caption{In-plane helical ordering in \BCAO, showing the stacking of quasi-ferromagnetic chains weakly coupled.}
\label{Figure-HelicalStructure}
\end{figure}

When the space dimension $D$ is lowered from three, to two (2D, planar system) and finally to one (1D, chain system), the magnetism displays more and more interesting and non-trivial features, as a result of the enhancement of both the thermal and quantum fluctuations. In the extreme cases, this leads to the lack of three-dimensional long-range ordering (LRO) down to T=0 K, the occurrence of spin-liquid states, and finally the emergence of unconventional spin dynamics. As established from numerous theoretical studies, both the ground state (GS) and the excited states of low-dimensional quantum magnets appear more and more exotic as the dimension of the spin-space $n$ increases ($n = 1$ for the Ising system, $n=2$ for the XY system and $n=3$ for the Heisenberg system), or the spin quantum number  $S$ decreases (from $S=\infty$ for the classical case, down to $S=1$ and $S=1/2$ for the extreme quantum spin). The nature of the ground state depends also strongly on the connectivity of the lattice (i.e. the number of next-nearest neighbor spins) and the type of spin-spin couplings which are involved: Ferromagnetic (F), antiferromagnetic (AF) or competing between first and second neighbor spins,
frustrating or not frustrating the spin lattice, at short range or at long range. As it is now well admitted, the largest effects are seen for the 1D antiferromagnetic Heisenberg (HAF) chain system, which indeed displays drastically different ground states (GS) and spin-excitation spectra, depending whether the spin value is half-integer ($S=1/2$, $3/2$, ...) or integer ($S=1$,$2$, ...)~\cite{1D-Haldane}. More precisely, for the latter Haldane predicted the existence of a non-magnetic $S=0$ singlet GS separated from the first triplet of excited states by a quantum energy gap $E_G \sim JS exp(-\pi S)$, J being the inter-spin coupling constant ($E_G \approx 0.41 J$ for $S=1$ from numerical calculations~\cite{HaldaneGap}). Conversely, for the $S=1/2$ HAF chain, the magnetic excitation spectrum should be a gapless continuum of magnetic excitations~\cite{1D-HAF}, but the introduction of frustrating AF second-neighbor interactions ($J1-J2$ model) leads to the opening of a gap above some critical ratio, by spontaneous dimerization of the spin system~\cite{Haldane-J1J2}. Similar effects have been predicted for the $S=1/2$ $p$-leg AF spin-ladder system, which indeed realize the cross-over between the $S=1/2$ HAF chain ($p=1$) and the $S=1/2$ HAF square plane ($p=\infty$). For p even-integer ($p=2$, $4$, ... ), the GS of the $S=1/2$ HAF p-leg spin-ladder system should again exhibit a non-magnetic $S=0$ singlet ground state, and a gapped triplet of first excited states, with a gap energy exponentially vanishing as $p \rightarrow \infty$ (with $E_G \approx 0.5 J$ for $p=2$)~\cite{2legSpinLadder}. At $D=2$, unlike the 1D case, the quantum fluctuations are less relevant than the thermal fluctuations, and the magnetism recovers a more classical behavior, with a phase transition at finite temperature toward LRO for the 2D Ising model and no phase transition down to 0 K for the 2D Heisenberg model, and for both, spin-wave-type excitations. As shown longtime ago by Kosterlitz and Thouless in their seminal paper~\cite{KT-Transition}, the 2D-XY ($D=2$, $n=2$) model is pathological: A phase transition occurs below a finite temperature $T_{KT} \sim (\frac{\pi}{2})JS^2$ (the famous Kosterlitz-Thouless transition temperature) from a disordered phase populated with unbind vortex and anti-vortex, toward a new topological ordered phase made of vortex-antivortex pairs, displaying an infinite susceptibility down to 0 K. The KT transition was further shown to be robust against in-plane anisotropy terms of 4-fold  or 6-fold symmetry~\cite{Jose-Kadanoff}, the latter condition being realized for the hexagonal lattice or the honeycomb lattice, build from two hexagonal sublattices. \\ 
Motivated by the recent discovery of very fascinating electrical properties in graphene~\cite{graphene}, and a renewed interest for studies of the magnetic properties of the honeycomb lattice~\cite{Iridate,alphaRuCl3}, we have reinvestigated the magnetic properties of the quasi-2D honeycomb-lattice planar systems of general chemical formula BaM${_{2}}$(X${_{4}}$)${_{2}}$ (M=Co,Ni, Fe; X=P, As). Among the various elements of this series, the cobaltite {\BCAO} (hereafter abbreviated as BCAO) exhibits very interesting and unusual magnetic properties (see, e.g., Ref.\cite{BCAO-deJongh} for a comprehensive review), currently not yet fully elucidated. BCAO crystallographic structure is a two Bravais-sublattice structure, described within the centrosymmetric trigonal (rhombohedral) space group $R\bar{3}$~\cite{BCAO-deJongh,Eymond-Durif69,EymondDurif69,BCAO-LPR83,Dordevic-structure}.
Within the hexagonal-cell representation, the lattice constants are $a=b=4.95$ \AA, $c=23.23$ {\AA}, and ${\beta}=120^{\circ}$ at $T \approx 1.5$ K. The crystallographic structure of BCAO can be viewed as a stacking of honeycomb-lattice layers of Co$^{2+}$ (electronic configuration $3d^{7}$, $L=3$ and $S=3/2$) ions (nearest neighbor in-plane Co-Co distance $\frac{a}{\sqrt{3}} \approx 2.85$ \AA), well separated along ${\bf{c}}$ by a distance $\frac{c}{3} \approx 7.74$ \AA (see Fig.~\ref{Figure-StructCristallo}). The high ratio ($\approx 2.72$) of the interlayer to in-plane nearest neighbor (n.n.) distances confers to BCAO a very pronounced quasi-2D magnetic character. From single-crystal susceptibility and magnetization measurements, it was shown that the magnetic moments are mainly located within the basal (${\bf{a}}$, ${\bf{b}}$) planes, BCAO being a very good planar system. The macroscopic magnetic properties of this material can be quantitatively analyzed by considering a model of strongly anisotropic $S=1/2$ effective spins (characteristic of a doublet ground state well separated from the first excited doublet state), interacting through the following XXZ spin Hamiltonian:~\cite{BCAO-deJongh,BCAO-LPR83,BCAO-suscep,BCAO-Excitations}
\begin{equation}
H=-{\sum}_{i,j} J_{ij}\left( S_{i}^{x}S_{j}^{x}+S_{i}^{y}S_{j}^{y}+\alpha_{z}S_{i}^{{z}}S_{j}^{z} \right) - 
       {\sum}_{i,\nu}g_{\nu}{\mu}_{B}S_{i}^{\nu}H_{\nu}  
\label{Eq-Hamiltonian}
\end{equation}
with effective spin-spin coupling constants $J \sim 30$ K (mostly ferromagnetic), an out-of-plane (OP) anisotropy parameter $\alpha_{z} \approx 0.4$, and anisotropic components of the gyromagnetic tensor, $g_{x} \approx g_{y} \approx 5$ and $g_{z} \approx 2.5$ (planar-type anisotropy). Only a very weak in-plane (IP) anisotropy could be detected from the magnetic susceptibility measurements on single crystal. 
The easy-planar and quasi-2D characters of magnetism in BCAO were corroborated by the observation of a rather strong $AT^{2}$ term (with $A \approx 28$ mJ/K$^3$/mole) in the low-temperature magnetic specific heat,~\cite{BCAO-chasp}, which a priori could be understood from the presence of a linear spin-wave (SW)  branch in the low-energy magnetic excitation spectrum. Comprehensive specific-heat and unpolarized neutron-diffraction measurements in zero-field have revealed a certain number of very peculiar features like, e.g., i) the occurrence of a sharp phase transition below 
$T_{N} \approx 5.35$ K ($kT_{N}/J \approx 0.16$), characterized by the incommensurate (IC) propagation vector ${\bf{k}}_{1}=(k_{x}, 0, k_{z})$, with $k_{x}=0.270{\pm}0.005$ and $k_{z} \approx-1.31$ (and equivalent wave vectors generated by trigonal symmetry, ${\bf{k}}_{2}=(-k_{x}, k_{x}, k_{z}$), and ${\bf{k}}_{3}=(0, -k_{x}, k_{z})$), ii) the absence of higher-order harmonics $n{\bf k}_{1}$ ($n$=$2$,$3$,$4$, ...), and iii) a remarkable step-like temperature dependence of the order parameter~\cite{BCAO-deJongh,BCAO-LPR83,BCAO-suscep,BCAO-Excitations}. All these results were qualitatively understood by assuming the in-plane helical structure shown in Fig.\ref{Figure-HelicalStructure}, described as a stacking of quasi-ferromagnetic (zigzag) pseudo-chains running along the {\bf {b}}-axis, with a phase angle $2{\pi}k_{x} \approx 96^{\circ}$ (close to $90^{\circ}$) between two adjacent pseudo chains~\cite{BCAO-suscep,BCAO-deJongh}. The phase angle between the two Bravais sublattices ${\phi_{12}} \equiv {\phi}  \approx  {83^{\circ}}$, is also close to $90^{\circ}$, implying weak effective inter-chain couplings. This is very likely the origin of the step-like staggered magnetization and the very peculiar field-temperature (H-T) phase diagram found in BCAO~\cite{BCAO-deJongh,BCAO-HT}. 
Under a magnetic field applied along $b$-axis, the spin system undergoes two successive first-order phase transitions, first toward an intermediate ferrimagnetic quasi-2D collinear phase at a critical field $H_{c1}(T \approx 0) \approx 0.33$ T, and second, toward the saturated paramagnetic state at a critical field $H_{c2}(T \approx 0) \approx 0.55$ T. The intermediate in-plane ferrimagnetic structure can be described as a
stacking of ferromagnetic chains parallel to the field direction, following a long-range ordered sequence ${\dots} {\uparrow} {\uparrow} {\downarrow} {\uparrow} {\uparrow} {\downarrow} {\dots}$ along the ${\bf{a}}^{*}$-direction. The ferrimagnetic ordering is characterized by a $1/3$-magnetization plateau between $H_{c1}$ and $H_{c2}$ and a planar propagation vector ${\bf{k}}_{Ferri}=(\frac{1}{3}, 0)$, with a completely random stacking (with probability $1/3$, corresponding to the three ${\uparrow}$, ${\uparrow}$, and ${\downarrow}$ moment possibilities) along the c-axis~\cite{BCAO-HT,BCAO-deJongh}.  Quite surprisingly, considering the rather strong IP exchange couplings involved in BCAO, the reduced critical fields $g_{x}{\mu}_{B}H_{c1}/J \approx 0.015$ and $g_{x}{\mu}_{B}H_{c2}/J \approx 0.020$ are very small, this implying that the helical, ferrimagnetic and saturated-paramagnetic structures should have very close magnetic energies. In other words, in BCAO the rotation of long chain-segments seems not costing much energy and low-energy magnetic defects can be easily created. This peculiarity explains the strong hysteresis and the metastable behavior observed at low T, especially when the magnetic field is decreased from above $H_{c2}$ down to zero (defined in the following as the  $0^{+}$ field)~\cite{BCAO-deJongh,BCAO-HT}.\\
The magnetic excitation spectrum of BCAO remains also quite intriguing \cite{BCAO-deJongh,BCAO-LPR83,BCAO-Excitations,BCAO-energy(T)}. Despite careful and extensive inelastic neutron scattering (INS) measurements performed both on thermal and cold neutron three-axis spectometers, no linear spin-wave (SW) branch emerging from the IC wave vector ${\bf{k}}_{1}$ could be detected (as it should have been, e.g., for a simple helimagnetic or helical structure), in spite of the strong $T^{2}$-term observed in the low-T magnetic specific heat. Contrary, the dispersion curves along the non-equivalent [1 0 0] and [1 1 0] directions exhibit a clear line of minima at wave vectors ${\bf{q}}=(0, 0, q_c)$ and a sharp spin-gap of energy $\Delta_{0} \approx 1.45$ meV, a behavior rather reminiscent of a quasi-2D gapped ferromagnetic mode. Unexpectedly, the simple SW theory for the honeycomb lattice described by the Hamiltonian (\ref{Eq-Hamiltonian}) with spin-spin interactions up to third neighbors failed to account for the magnetic excitation spectra of the zero-field ground-state and of the intermediate ferrimagnetic phase. In contrast, the simple SW theory is quite successful to reproduce the dispersion of magnetic excitations in the saturated-paramagnetic phase (described by the propagation vector ${\bf k}=0$), in magnetic fields applied along the $b$-direction~\cite{BCAO-deJongh,BCAO-Excitations}.  For $H>H_{c2}$, a good agreement between the experimental and calculated SW dispersion curves (both acoustic and optical) could be achieved by taking coupling parameters between first-, second- and third-neighbor spins (see Fig.\ref{Figure-StructCristallo}), $J_{1} \approx 38$ K, $J_{2} \approx 1.3$ K ($J_{2}/J_{1} \approx 0.03$) and $J_{3} \approx -10$ K ($J_{3}/J_{1} \approx -0.26$), an out-of-plane (OP) anisotropy term $\alpha_{z} \approx 0.37$, and a very small IP anisotropy ratios${\mid}\frac{J_{n}^{x}-J_{n}^{y}}{J_{n}}{\mid} < 0.01$ (n=1, 2 and 3). The latter, much too small, cannot be held responsible for the opening of the 1.45-meV  gap at ${\bf{q}}=0$, which still remained very puzzling. Furthermore, with the above exchange-coupling parameters, the ground-state structure should have rather been ferromagnetic than helimagnetic. Obviously, there exist some inconsistencies among the published data, whose solution deserve further experimental investigations. \\
In order to clarify the nature of  the magnetic ordering in BCAO (especially the role played by the frustration) and understand its puzzling spin dynamics, we have performed a comprehensive  investigation of elastic and inelastic magnetic contributions in this material by means of the spherical-neutron-polarimetry (SNP) technique.\

\section{Methodology}
\label{SNP}

The neutron-scattering technique, due to the neutron specificities (among other, because it's a massive and neutral particle, bearing a spin $1/2$), is an invaluable technique for probing the magnetic properties in bulk materials. The bases of the technique are described in several seminal textbooks ~\cite{LPA-textbook,Lovesey87,Balcar-Lovesey89}, emphasizing the relevance of the polarized-neutron scattering and the longitudinal polarization analysis (LPA) for the determination of magnetic structures and magnetic excitation spectra. By principle, the LPA allows to only recover the projection of the final polarization vector, ${\bf{P}}$, onto the incident polarization vector ${\bf{P}}_{0}$, leading to an important loss of information. The recent availability of a new generation of diffractometers and three-axis spectrometers (TAS) providing high flux of polarized neutrons, in conjunction with the use of more sophisticated polarization-analysis methods, capable of determining independently the three components of the neutron polarization vector after scattering, has open a new field of investigation of materials exhibiting non-conventional magnetism~\cite{SNP-GEN}. The technique, referenced in the literature as vectorial neutron polarimetry (VNP) or spherical-neutron-polarimetry (SNP), is based on the use of the cryogenic polarization-analysis device CRYOPAD, invented more than fifteen years ago by F. Tasset~\cite{Tasset-SNP-89} for the diffraction, and recently optimized for the inelastic neutron scattering (INS)~\cite{LPR-SNP-04,ELB-SNP-05,JAEA-SNP-05,ELB-SNP-07}. The principle of this device relies on the use of the combination of two pairs of magnetic fields (nutation and precession) decoupled by niobium-based superconducting shields in order to independently control the incident polarization vector (${\bf{P}}_{0}$) and analyze independently the three components of the final neutron polarization vector ($\bf{P}$). By principle, the sample is located in a zero-field area (residual magnetic field smaller than $2$ mG), this allowing to keep intact the neutron polarization after scattering, until its analysis. For a given scattering vector ${\bf{Q}}$, by measuring the three components of the final polarization vector for three different orientations of the incident polarization, one is able to determine a 3x3 matrix, called the polarization matrix $P_{\alpha \beta}(\bf{Q})$ (${\alpha ,\beta=x,y,z}$), which contains all the information on the various structural and magnetic cross-sections, as we shall see later.\\
The general expressions giving the polarization vector $\bf{P}$ of a scattered neutron beam as a function of the incident polarization vector  ${\bf{P}}_{0}$ were derived long ago in two seminal papers~\cite{Maleyev-SNP-63,Blume-SNP-63}. The reader interested by the SNP can find all the necessary information in two recently published textbook chapters~\cite{JB-SNP-Chatterji,LPR-SNP-Chatterji}. Basically, for a given scattering vector $\bf{Q}$, the SNP method gives access to the polarization matrix elements  $P_{\alpha \beta}(\bf{Q})$ for an incident polarization direction  ${\alpha}$ and a polarization-analysis direction  ${\beta}$ (${\alpha ,\beta=x,y,z}$). In practice, this can be achieved from the measurement of the two scattering cross-sections associated with neutron spin states  ${|+\rangle}$ ($\sigma_{\alpha \beta}^{+}(\bf{Q})$) and ${|-\rangle}$ 
($\sigma_{\alpha \beta}^{-}({\bf{Q}})$), according to the relation: 
$P_{\alpha \beta}({\bf{Q}})=\frac{\sigma_{\alpha \beta}^{+}(\bf{Q})-\sigma_{\alpha \beta}^{-}(\bf{Q})}
 {\sigma_{\alpha \beta}^{+}(\bf{Q})+\sigma_{\alpha \beta}^{-}(\bf{Q})}$.
The coordinate frame used in this paper is defined as follows:  ${\bf{x}}\parallel{ \bf{Q}}$,  ${\bf{y}}\perp {\bf{Q}}$
in the scattering plane (${\bf{k}}_i$, ${\bf{k}}_f$), and $\bf{z}$ vertical, perpendicular to the scattering plane. Subject to the existence of an axial vector in the problem,~\cite{Maleyev-Chiral} ${\bf{P}}$ is in general not collinear to ${\bf{P}}_{0}$. The neutron polarization may undergo a small rotation, and the polarization matrix may have off-diagonal elements, in addition to the usual diagonal ones. As shown in Refs.~\cite{Maleyev-SNP-63} and \cite{Blume-SNP-63}, for the coordinate frame previously defined, in the most general case the neutron-polarization vector ${\bf{P}}$ will depend on the combination of up to nine different correlation functions, which involve the nuclear ($N_{Q}$) and magnetic ($M_{\perp Q}^{\alpha}$) amplitude operators ($\alpha$=($x$, $y$, $z$)), namely:
\begin{eqnarray}
N &=& \langle N_{Q} N_{Q}^{\dagger}\rangle_{\omega} \nonumber \\
M_{yy} &=& \langle M_{\perp Q}^{y} M_{\perp Q}^{y\dagger}\rangle_{\omega} \nonumber \\ 
M_{zz} &=& \langle M_{\perp Q}^{z} M_{\perp Q}^{z\dagger}\rangle_{\omega} \nonumber \\
M_{ch} &=& i(\langle M_{\perp Q}^{y} M_{\perp Q}^{z\dagger}\rangle_{\omega}-\langle M_{\perp Q}^{z} M_{\perp Q}^{y\dagger}\rangle_{\omega}) \nonumber \\
M_{yz}^{+} &=& \langle M_{\perp Q}^{y} M_{\perp Q}^{z\dagger}\rangle_{\omega}+\langle M_{\perp Q}^{z} M_{\perp Q}^{y\dagger}\rangle_{\omega} \nonumber \\
R_{y} &=& \langle N_{Q} M_{\perp Q}^{y\dagger}\rangle_{\omega}+\langle N_{Q}^{\dagger} M_{\perp Q}^{y}\rangle_{\omega} \nonumber \\
I_{y} &=& i(\langle N_{Q} M_{\perp Q}^{y\dagger}\rangle_{\omega}-\langle N_{Q}^{\dagger} M_{\perp Q}^{y}\rangle_{\omega}) \nonumber \\
R_{z} &=& \langle N_{Q} M_{\perp Q}^{z\dagger}\rangle_{\omega}+\langle N_{Q}^{\dagger} M_{\perp Q}^{z}\rangle_{\omega} \nonumber \\
I_{z} &=& i(\langle N_{Q} M_{\perp Q}^{z\dagger}\rangle_{\omega}-\langle N_{Q}^{\dagger} M_{\perp Q}^{z}\rangle_{\omega}) \nonumber 
\nonumber 
\end{eqnarray}
in which $N_{Q} =N^{1/2} \sum\limits_{j}b_{j}e ^{i{\bf Q}\cdot{\bf R}_{j}}$ (the nuclear scattering amplitude scalar operator), and 
${\bf M}_{\bot Q} = r_{0}N^{1/2} \sum\limits_{j}\left[ {\bf M}_{j} -({\bf Q} \cdot {\bf M}_{j} ) \cdot {\bf Q}/Q^{2}\right] e^{i{\bf Q} \cdot {\bf R}_{j}}$ (the magnetic scattering amplitude vectorial operator),
with $b_{j}$, the scattering length associated with nucleus $j$ ($b_{Co} \approx 0.25{\times}10^{-12}$ cm and $b_{O} \approx 0.58{\times}10^{-12}$ cm), $r_{0} \approx 0.54 \times 10^{-12}$ cm, and ${\bf M}_{j} = {\bf s}_{j} -\frac{i}{\hbar }\frac{{\bf Q}\times {\bf p}_{j}}{{Q}^{2}}$ 
depends on both the electron spin (${\bf s}_{j}$) and electron momentum (${\bf p}_{j}$)~\cite{Lovesey87,Balcar-Lovesey89,Trammell53}.
In the above expressions, $N$, $M_{yy}$, $M_{zz}$ are the standard purely nuclear and magnetic terms,  $M_{ch}$ is the chiral term, associated with the antisymmetric purely magnetic cross-correlation functions, and $M_{yz}^{+}$ corresponds to the symmetric purely magnetic cross-correlation functions. The last four terms are the symmetric ($R_{y}$ and $R_{z}$ ) and anti-symmetric ($I_{y}$ and $I_{z}$) nuclear-magnetic interference (NMI) terms, mixing the nuclear and magnetic components. In all these relations, $\langle A_{Q}B_{Q}^{\dagger} \rangle_\omega$ represents the Fourier transform on space and time of the pair
correlation function $\langle A({\bf R}_{n},t) B^{\dagger}({\bf 0},0) \rangle$:
\begin{equation}
{\langle}A_{Q}B_{Q}^{\dagger}{\rangle}_{\omega} =\frac{1}{2{\pi}} \int \limits_{-\infty}^{+\infty} dt \, e^{i \omega t}
\, \sum\limits_{n} {\langle}A({\bf R}_{n},t) B^{\dagger}({\bf 0},0){\rangle} e^{i{\bf Q}{\cdot}{\bf R}_{n}}  \nonumber
\label{Eq-correl}
\nonumber
\end{equation}
 and $M_{\alpha \beta}=\langle M_{\perp Q}^{\alpha} M_{\perp Q}^{\beta \dagger}\rangle_{\omega}$ ($\alpha, \beta=y, z$) are the pure magnetic cross-correlation functions, which involve only the magnetic components perpendicular to the scattering vector {\bf{Q}}. 
In principle, all these terms can be redundantly determined from the measurement of the polarization matrices, by applying the Blume-Maleyev formalism. However, the off-diagonal terms, which involve directly the various domain populations and the (t, -t) time-reversal symmetry, may easily vanish by symmetry.
For the diffraction case, the expressions giving the various structure-factor components are very similar and obtained just by replacing the various operators by the corresponding vector components, and the correlation functions ${\langle}A_{Q}B_{Q}^{\dagger}{\rangle}_{\omega}$ by simple products $A_{Q}(B_{Q})^*$, where $A_{Q}$ and $B_{Q}$ are two structure-factor components, and $(B_{Q})^{*}$ is the complex conjugate of $B_{Q}$. In the following, we will define the various elastic  magnetic structure-factor components as $M_{\alpha \beta}=M_{\alpha} (M_{\beta})^{*}$ ($\alpha, \beta=y, z$), where $M_{\alpha}$ is the Fourier transform of the $\alpha$ component of the moment distribution. \\
The correlation functions $M_{ch}$, $R_{y}$, and $R_{z}$ can be determined before anything else, just by measuring the
polarization creation after scattering from an initially unpolarized beam (e.g., by using a configuration with a graphite or copper
monochromator, and an Heusler analyzer, or the other way around), and applying the relations:
\begin{eqnarray}
P_{0x} & = &  \frac{M_{ch}}{N+\sigma_{M}}  \label{Eq-P0x}\\
P_{0y} & = &  \frac{R_{y}}{N+\sigma_{M}}   \label{Eq-P0y}\\
P_{0z} & = &  \frac{R_{z}}{N+\sigma_{M}}    \label{Eq-P0z}
\end{eqnarray}
in which ${\sigma_{M}}=M_{yy}+M_{zz}$ reflects the total magnetic cross-section. 
For a purely magnetic contribution (i.e., by neglecting the N and NM terms), the following relations can be derived for the
diagonal and the off-diagonal components:
\begin{eqnarray}
P_{xx} &\approx& -\frac{P_{0}\sigma _{M}-M_{ch}}{\sigma _{M}-P_{0}M_{ch}}  \label{Eq-Pxx-Mag}\\
P_{yy} &=&-P_{zz} \approx \frac{M_{yy}-M_{zz}}{\sigma _{M} } P_{0}  \label{Eq-Pyy-Mag}\\
P_{xy} &\approx& P_{xz} \approx 0 \label{Eq-Pxy-Mag} \\
P_{yx} &\approx& P_{zx}\approx  \frac{M_{ch}}{\sigma _{M} }  \label{Eq-Pyx-zx-Mag}\\
P_{yz} &\approx& P_{zy}\approx \frac{M^{+}_{yz}}{\sigma _{M} } P_{0}  \label{Eq-Pyz-Mag} 
\end{eqnarray}
If a structural component $N$ is superimposed to the magnetic ones, in the absence of chiral and NMI terms, the diagonal elements $P_{xx}$, $P_{yy}$, and $P_{zz}$ are now given by the following relations: 
\begin{eqnarray}
P_{xx} &\approx& \frac{N-\sigma _{M}}{N+\sigma _{M} } P_{0}  \label{Eq-Pxx-N&M}\\
P_{yy} & \approx& \frac{N+M_{yy}-M_{zz}}{N+\sigma _{M} } P_{0}  \label{Eq-Pyy-N&M}\\
P_{zz} & \approx& \frac{N+M_{zz}-M_{yy}}{N+\sigma _{M} } P_{0}  \label{Eq-Pzz-N&M}
\end{eqnarray}
implying that $P_{yy} \neq -P_{zz}$.
In case of finite NMI and chiral terms, the off-diagonal elements are given by the following general expressions:
\begin{eqnarray}
P_{xy} &\approx& -\frac{I_{z}P_{0}}{N+\sigma _{M}+R_{y}P_{0}}  \label{Eq-Pxy-NMI}\\
P_{xz} &\approx& \frac{I_{y}P_{0}}{N+\sigma _{M}+R_{z}P_{0}}   \label{Eq-Pxz-NMI}\\
P_{yx} &\approx& \frac{M_{ch}+I_{z}P_{0}}{N+\sigma _{M}+R_{y}P_{0}}  \label{Eq-Pyx-NMI}\\
P_{zx} &\approx& \frac{M_{ch}-I_{y}P_{0}}{N+\sigma _{M}+R_{z}P_{0}}   \label{Eq-Pzx-NMI}\\
P_{yz} &\approx& \frac{R_{z}+M^{+}_{yz} P_{0}}{N+\sigma _{M}+R_{y}P_{0}}  \label{Eq-Pyz-NMI}\\ 
P_{zy} &\approx& \frac{R_{y}+M^{+}_{yz} P_{0}}{N+\sigma _{M}+R_{z}P_{0}}  \label{Eq-Pzy-NMI} 
\end{eqnarray}
which show that $P_{zx} \neq P_{yx}$ and $P_{yz} \neq P_{zy}$ in the most general case. Relations~(\ref{Eq-P0x})-(\ref{Eq-Pzy-NMI}) will be used later for the quantitative analysis of our SNP data in BCAO.\\
In most of cases, the intrinsic accuracy of the off-diagonal matrix elements is in the range $0.01-0.03$. However, some matrix elements (e.g., those involving the anti-symmetric correlation function $I_{z}$ or the symmetric correlation functions $M_{yz}^{+}$ and $R_{z}$) can be determined with improved accuracy (below $0.01$) by considering the anti-symmetric combination $P_{\alpha \beta }^{a}({\bf{Q}})=\frac{P_{\alpha \beta}({\bf{Q}})-P_{\alpha \beta}(-{\bf{Q}})}{2}=-P_{\alpha \beta}^{a}(-{\bf{Q}})$, which cancels at first order the systematic errors, expected to be invariant in a rotation of the sample by $180^{\circ}$. This trick has been used in our experimental determination of some off-diagonal elements of polarization matrices in BCAO. Thus, the SNP with CRYOPAD makes possible accurate investigations of magnetic, structural and hybrid (magneto-structural) correlation functions, through the measurements of polarization matrices as a function of the scattering vector (${\bf{Q}}$) and the energy transfer (${\hbar}{\omega}\equiv{\Delta}E$).\

\section{Experimental}
\label{experimental}

\begin{figure}
\centering
\includegraphics[width=10cm]{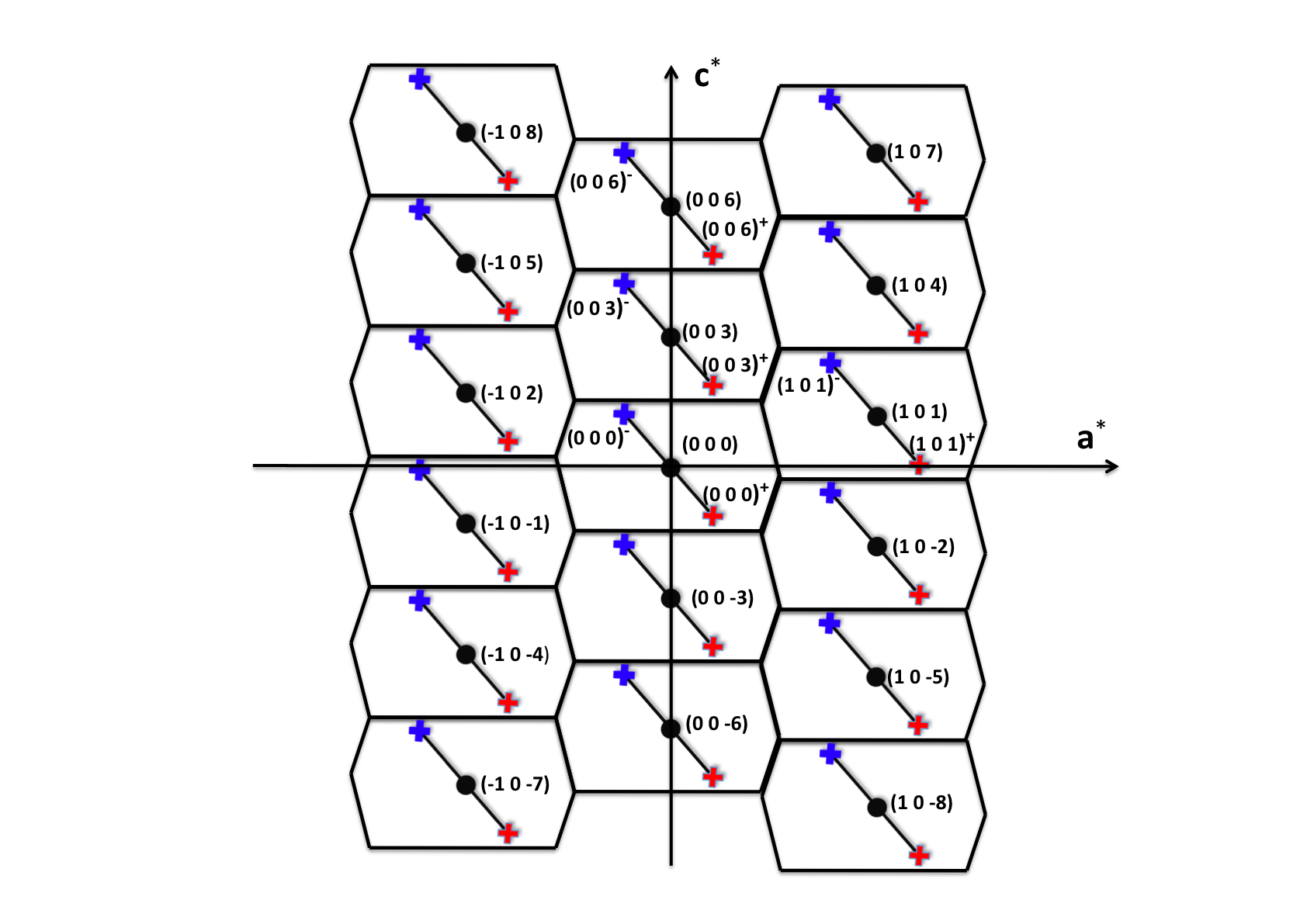}
\caption{Sketch of the (${\bf{a}}^{*}$, {${\bf{c}}^{*}$}) reciprocal-lattice plane, showing the locations of some structural Bragg reflections ${\bf{H}_{N}}=(h_{N},  0,  l_{N})$ (closed circles) and their associated magnetic satellites $(h_{N},  0,  l_{N})^{\pm}$ (crosses). The polygons around the structural Bragg spots represent the various Brillouin zones.}
\label{Figure-ReciprocalSpace}
\end{figure}

The SNP experiments on BCAO have been mainly performed on TAS IN22, high-flux instrument with polarized-neutron capabilities installed at the end position of the H25, m=2 supermirror guide, at the Institut Laue Langevin (ILL), Grenoble. The unpolarized INS experiments were performed by using pyrolithic graphite (PG) (002) monochromator and analyzer, at fixed final neutron wave vectors, $k_{f}=1.64$ \AA $^{-1}$, $1.97$ \AA$^{-1}$, and $2.662$ \AA$^{-1}$. The polarized-neutron configuration was the following: Variable vertical focusing Heusler-[111] monochromator; Two sets of Heusler-[111] analyzers were used: one with no vertical focusing
and variable horizontal focusing, and another with fixed vertical focusing (optimized for $k_{f}=2.662$ \AA$^{-1}$) and variable horizontal focusing. The measurements were performed at three different final neutron wave vectors $k_{f}=1.97$ \AA$^{-1}$, $2.662$ \AA$^{-1}$, and $3.84$ \AA $^{-1}$. In all cases, a 5-cm long PG filter was placed on the scattered beam in order to minimize the higher-order contamination (especially that at $2k_{f}$). 
For these experiments, we have mainly used CRYOPAD. Some measurements were performed by using the more classical Helmholtz-coils set-up. The flipping of the neutron polarization was performed by reversing the nutation field (CRYOPAD
configuration) or by using a Mezei-type flipper (Helmholtz-coils configuration), both located on the scattered-beam side. Depending on the contribution (elastic or inelastic) under investigation and the type of analyzer used, the flipping ratio $\rho_{F}$ ranged between 15 and 27. The crystal, shaped as a platelet of dimensions 14x14x1 mm$^{3}$ (V $\approx 0.2$ cm$^{3}$) was mounted on the cold finger of a standard ILL-type Orange cryostat and aligned with the {\bf{b}}-axis perpendicular to the scattering plane, in order to survey scattering vectors ($Q_{a}, 0, Q_{c}$).   The sample was an as-grown untwined ({\bf{c}}/-{\bf{c}}) single crystal, as documented by the very small intensity of the forbidden $(1, 0, -1)$ reflection, while the allowed $(1,0, 1)$ reflection is very strong. The corresponding intensity ratio, 20/13000, leads us to conclude on a nearly (99.9\%) untwined single crystal.
In the following, the scattering vectors ${\bf{Q}}=(h_{N}, 0, l_{N}){\pm}\bf{k}$ associated with the magnetic satellites of the structural reflection $(h_{N}, 0, l_{N})$ will be more compactly labeled $(h_{N}, 0, l_{N})^{\pm}$ (where $h_N$ and $l_N$ are integers verifying the trigonal extinction rule $-h_{N}+l_{N}=3n$, $n$ integer). A sketch of the (${\bf{a}}^*$, ${\bf{c}}^*$) reciprocal-lattice scattering plane  is shown in Fig.~\ref{Figure-ReciprocalSpace}. \

\section{Results}
\label{Results}

\begin{figure}
\centering
\includegraphics[width=13cm]{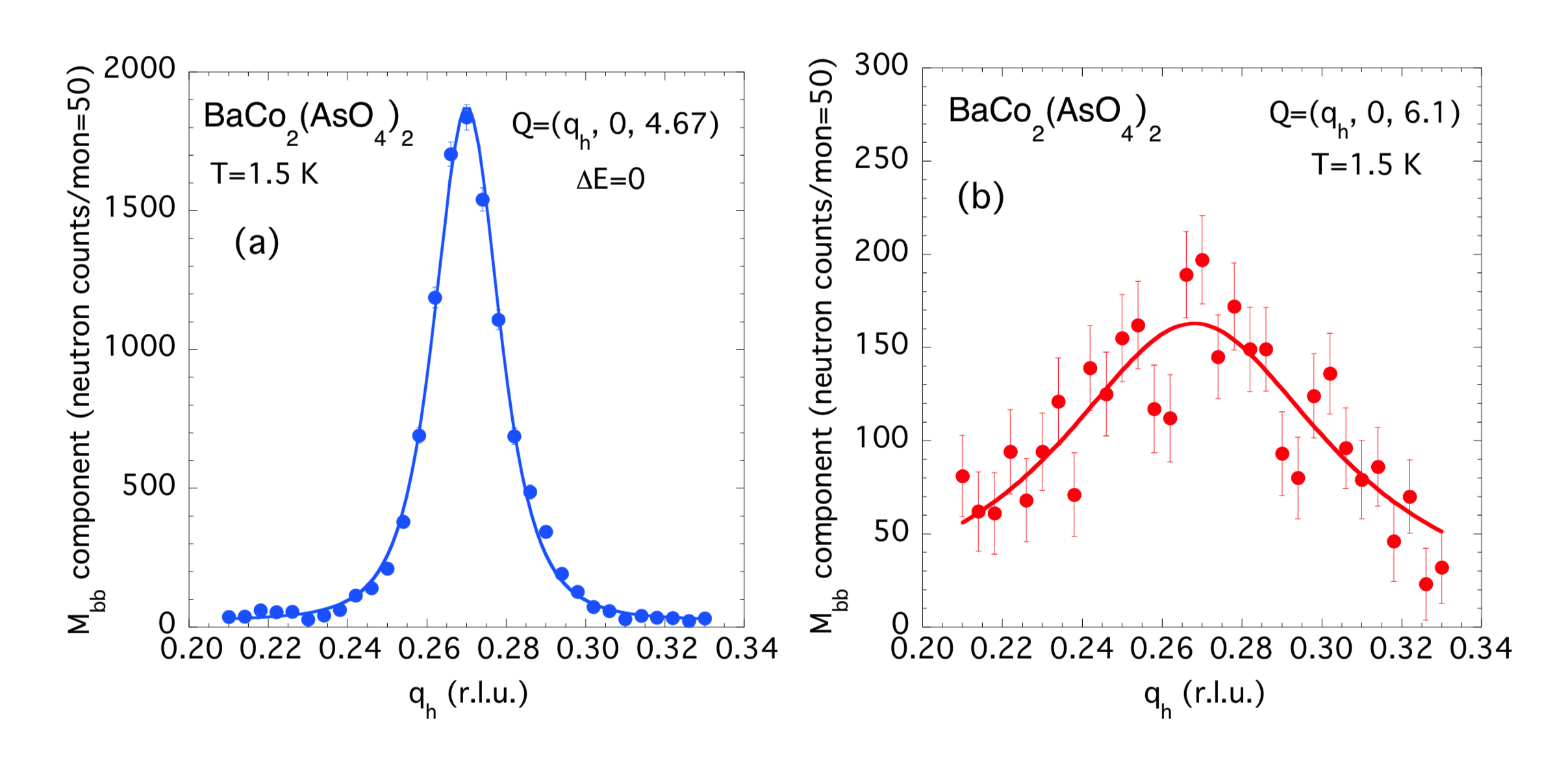}
\caption{Elastic scans along {\bf{a}}$^{*}$ at T=1.5 K, (a) across the magnetic zone center at {\bf{Q}}=(0.27, 0, 4.67)=(0, 0, 6)$^{+}$ and (b), close to the magnetic zone boundary at ${\bf{Q}}=(0.27, 0, 6.1)$ (b),  showing that the magnetic scattering extends over the entire Brillouin zone along ${\bf{c}}$.
The solid lines are fit to Lorentzian functions as described in the text.}
\label{Figure-ScanElastic}
\end{figure}

\begin{figure}
\centering
\includegraphics[width=13cm]{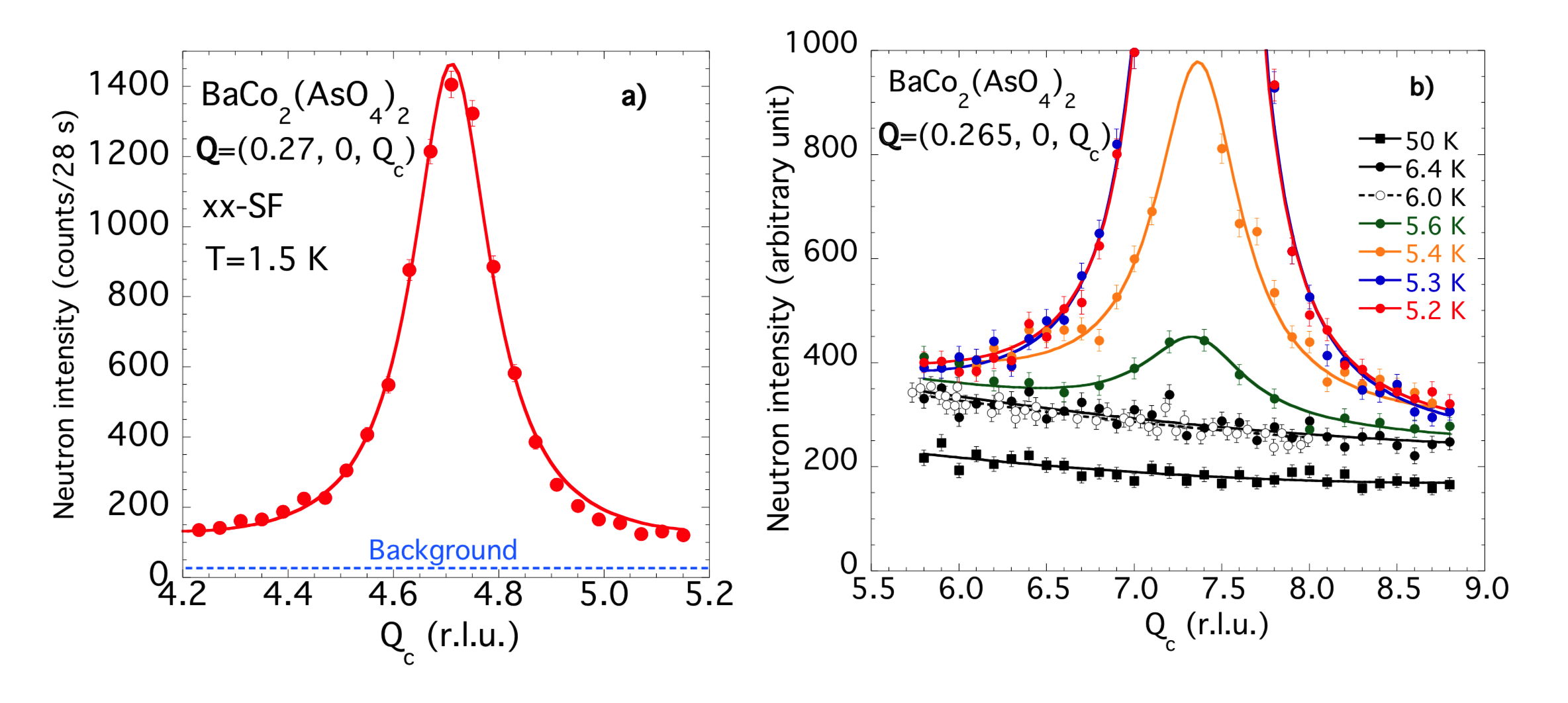}
\caption{a) Spin-flip elastic scan along {\bf{c}}$^{*}$ across the magnetic zone center $(0, 0, 6)^{+}$, performed with the polarization applied and analyzed parallel to the scattering vector (xx configuration). The solid line is a fit to a Lorentzian function, as described in the text. b) Unpolarized elastic scan along ${\bf{c}}^{*}$ across the magnetic zone center $(0, 0, 9)^{+}$, at several temperatures located from both sides of $T_{N} \approx 5.35$ K. The solid and dashed lines are guide to the eye.}
\label{Figure-Scan-Qc}
\end{figure}

\begin{figure}
\centering
\includegraphics[width=13cm]{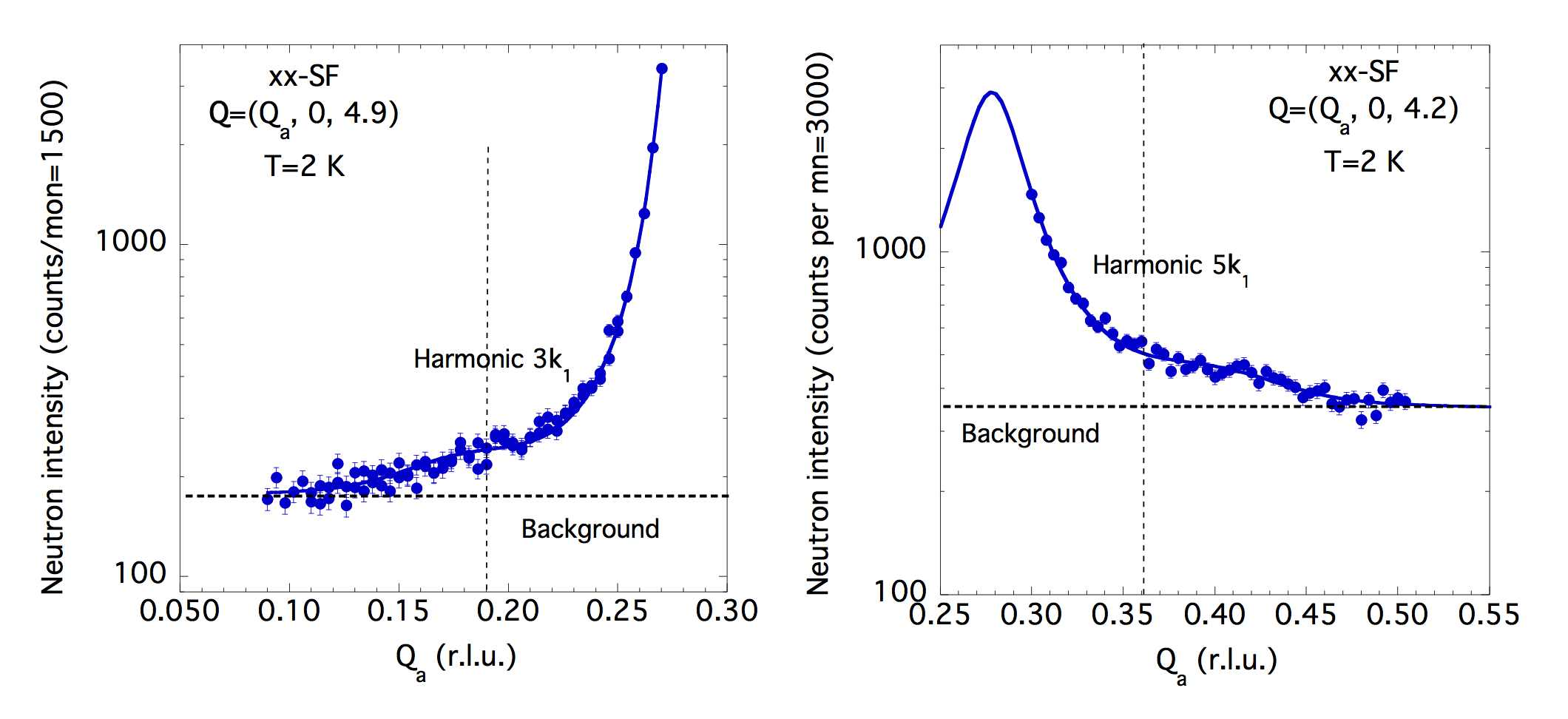}
\caption{xx-SF scans around the positions of the third harmonic at ${\bf{Q}}=(0.19, 0, 4.9)$ (left panel) and fifth harmonic at ${\bf{Q}}=(0.36, 0, 4.2)$ (right panel). The solid lines are guide to the eye.}
\label{Figure-Harmonics}
\end{figure}

\begin{figure}
\centering
\includegraphics[width=8cm]{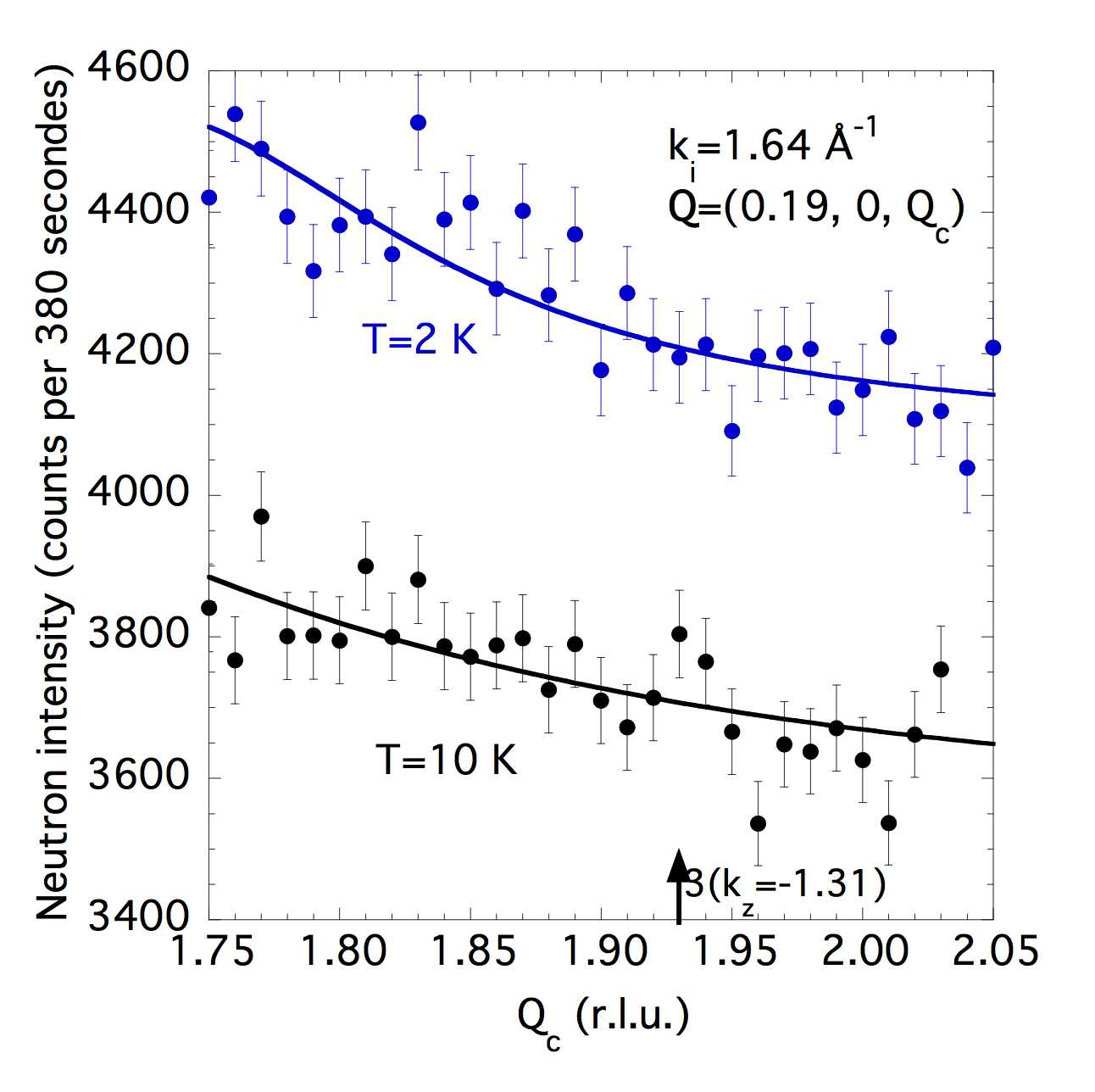}
\caption{Unpolarized scans at $k_{i}=1.64 \AA^{-1}$ around the position of the third harmonic at ${\bf{Q}}=(0.19, 0, 1.9)$ below ($T=2$ K) and above ($T=10$ K) $T_{N}$. The solid lines are guide to the eye.}
\label{Figure-Harmonics-Qc}
\end{figure}

\begin{figure}
\centering
\includegraphics[width=9cm]{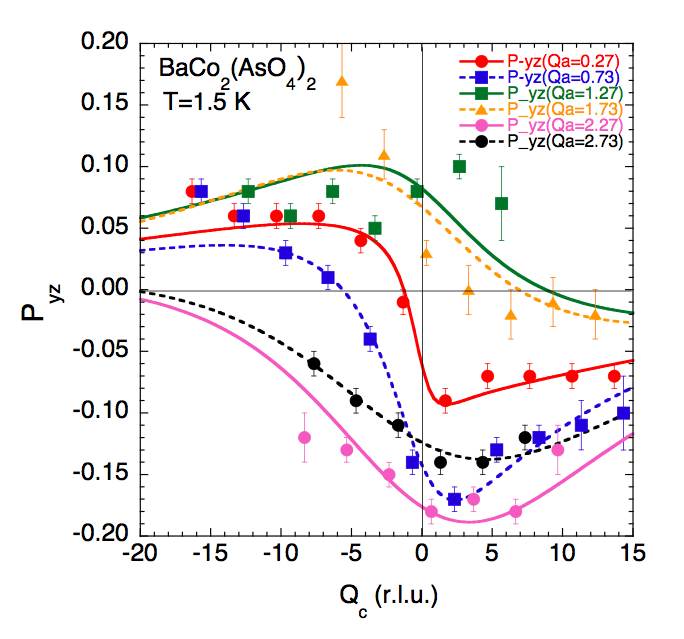}
\caption{$Q_{c}$-dependence of the $P_{yz}$ term at $T=1.5$ K, for several values of $Q_{a}$. The various lines are calculated as described in the text.}
\label{Figure-Pyz(Qc)}
\end{figure}

\begin{figure}
\centering
\includegraphics[width=8cm]{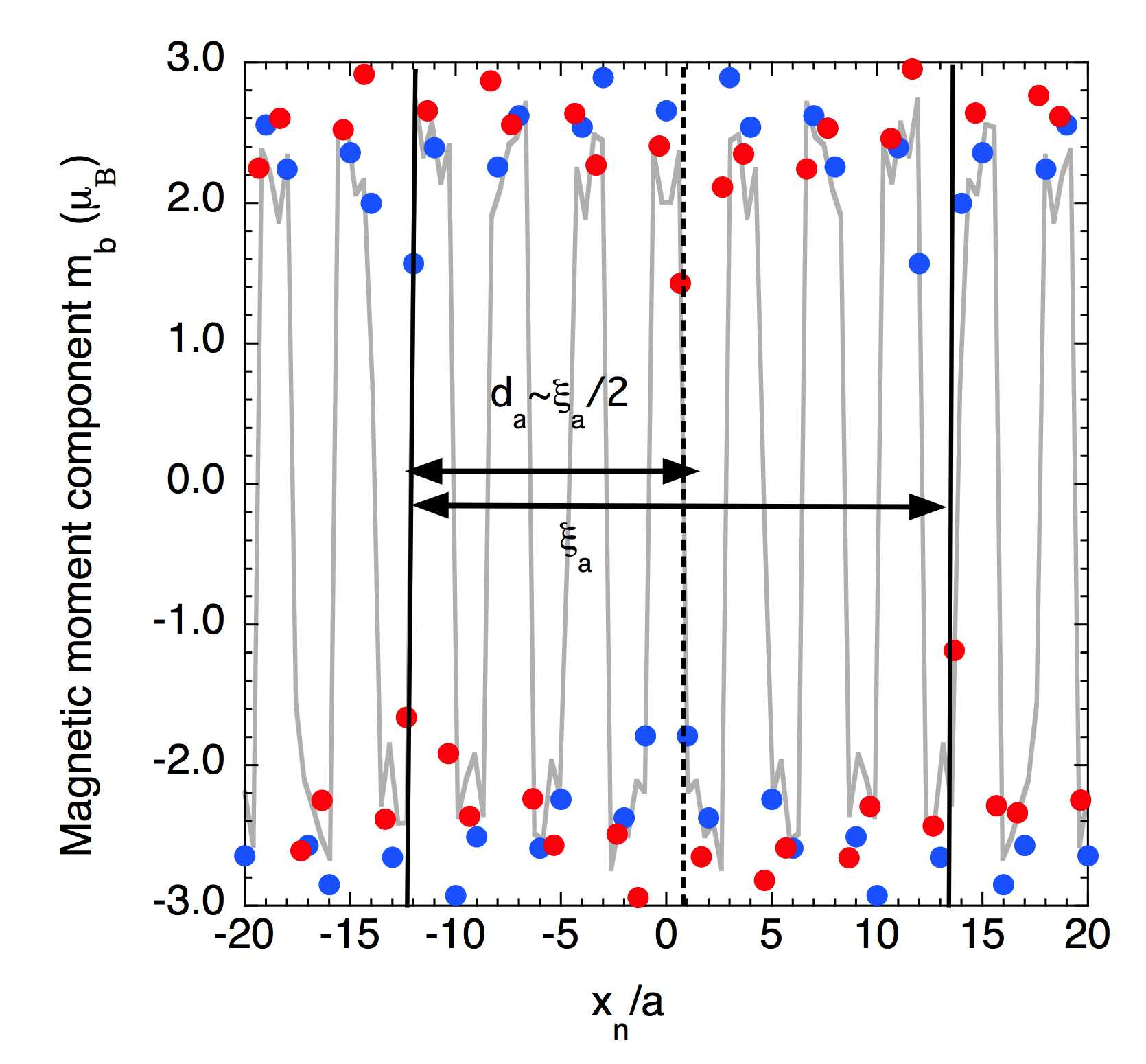}
\caption{Magnetic moment component along ${\bf{b}}$ as a function of the distance along the direction  ${\bf{a}}$ (calculated with the parameters resulting from the fit of components $P_{yz}(Q_{a}, Q_{c})$ and $P_{zy}(Q_{a}, Q_{c})$), showing the existence of more or less regular defects in the ${\dots} \uparrow \uparrow \downarrow \downarrow \uparrow \uparrow {\dots}$ sequence, separated by a distance $d_{a} \approx 13a \sim \frac{\xi_{a}}{2}$. Red closed symbols: Bravais sublattice 1; Blue closed symbols: Bravais sublattice 2.}
\label{BCAO-modulation}
\end{figure}

\begin{figure}
\centering
\includegraphics[width=8cm]{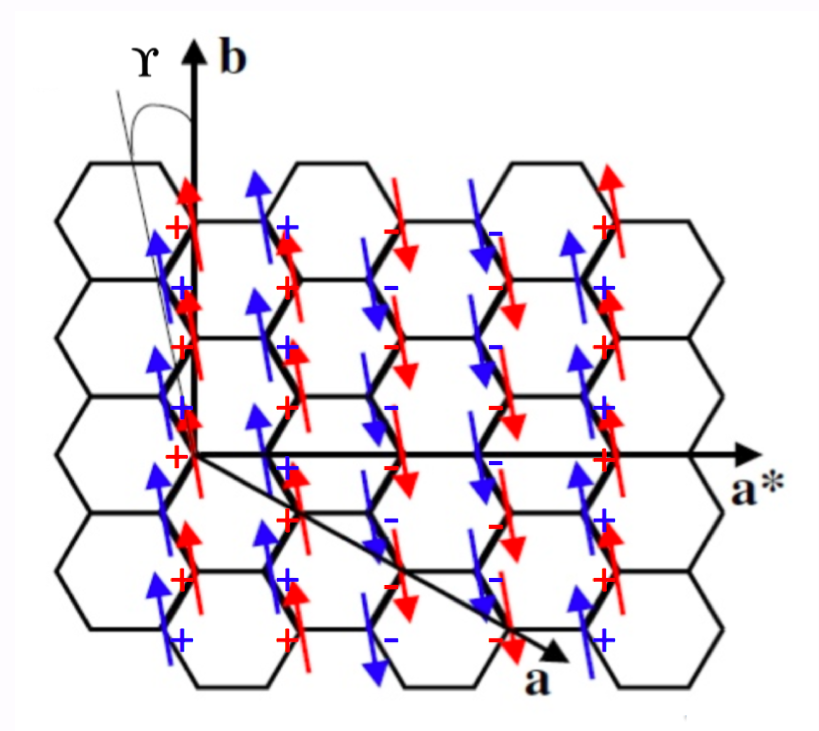}
\caption{(Squared-up collinear magnetic structure of {\BCAO}. In-plane (${\textcolor{black}{\uparrow}}$ and ${\textcolor{black}{\downarrow}}$) and out-of-plane ($+$ and $-$) magnetic ordering. $\gamma$ is the tilt angle w.r.t. the b-axis.}
\label{Figure-StructMagnCollinear}
\end{figure}

\begin{figure}
\centering
\includegraphics[width=13cm]{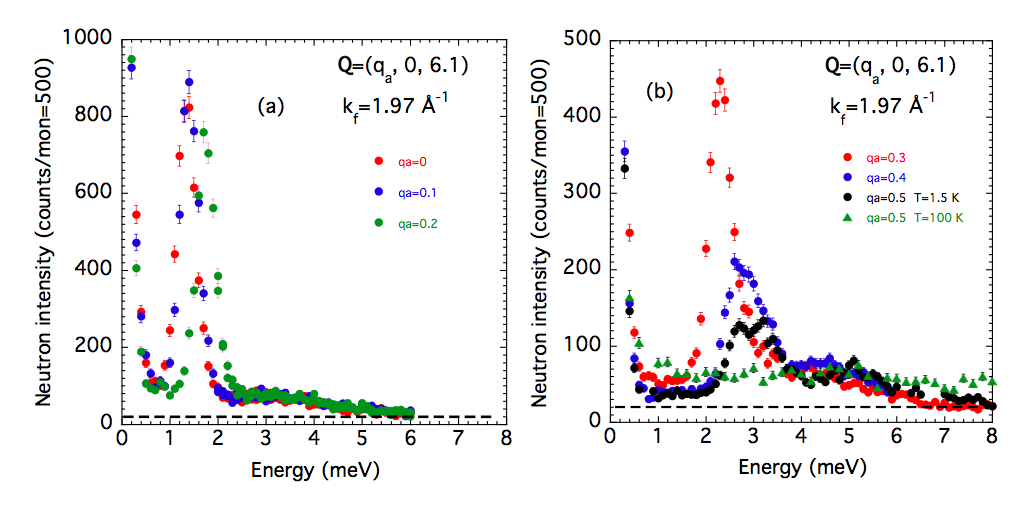}
\caption{Constant-Q scans  in {\BCAO} at $T=1.5$ K as a function of the reduced wave vector $q_{a}$, showing the dispersion of magnetic excitations  along the {\bf{a}}$^*$ direction. (a): $q_{a}=0$, $0.1$ and $0.2$ r.l.u.; (b): $q_{a}=0.3$, $0.4$, $0.5$ r.l.u. ($T=1.5$ K) and $q_{a}=0.5$ r.l.u. ($T=100$ K).}
\label{BCAO-MagnonGroup}
\end{figure}

\begin{figure}
\centering
\includegraphics[width=13cm]{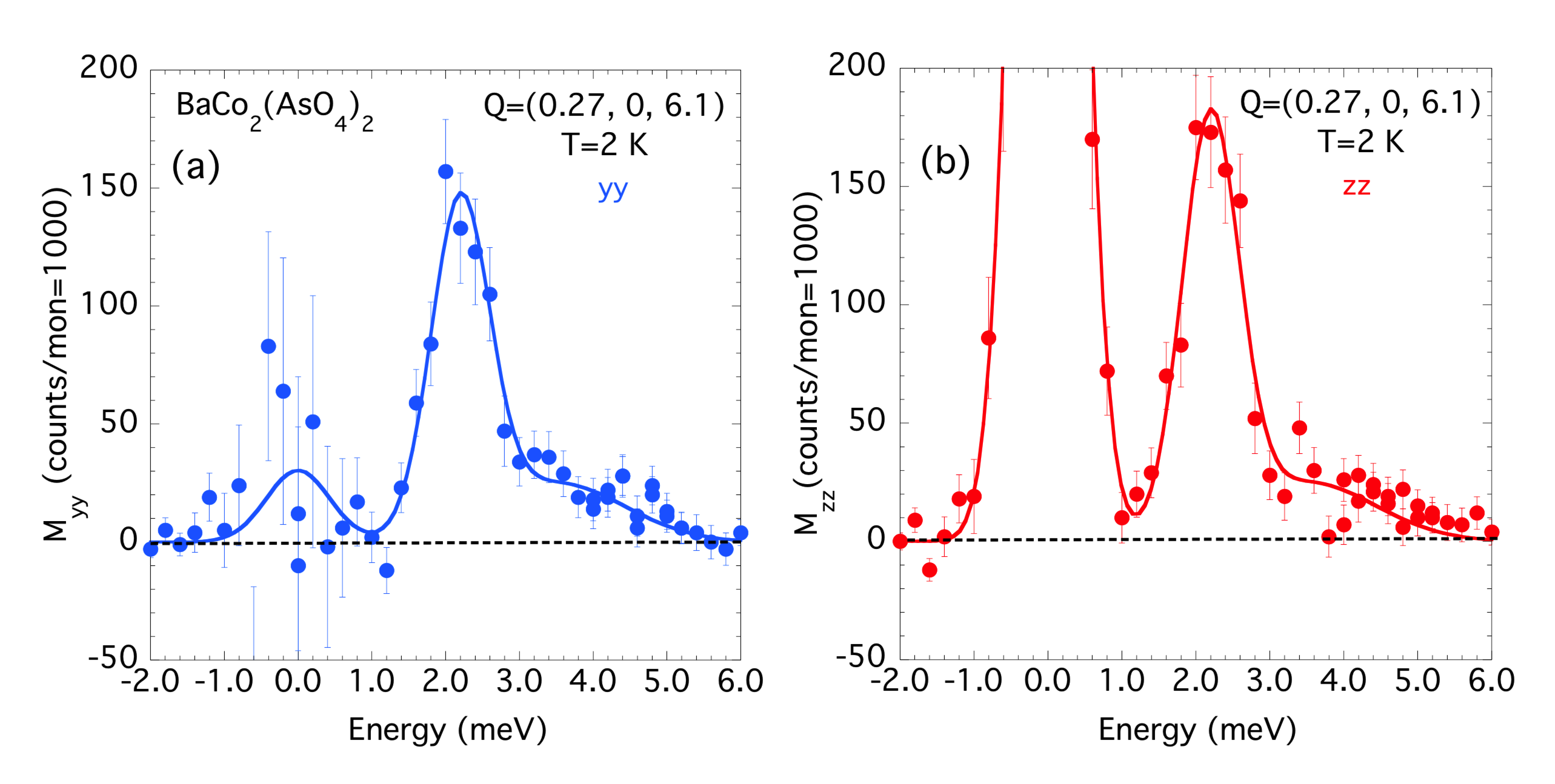}
\caption{Energy-dependence of the pure-magnetic components
$M_{yy}$ (a) and $M_{zz}$ (b), at the scattering vector ${\textbf{Q}}=(0.27, 0, 6.1)$, showing the almost isotropic character of magnetic excitations and the strong anisotropy of the elastic magnetic contributions. The solid lines are fit to a multi-Gaussian functional, as described in the text.}
\label{Figure-MyMz0.27}
\end{figure}

\begin{figure}
\centering
\includegraphics[width=13cm]{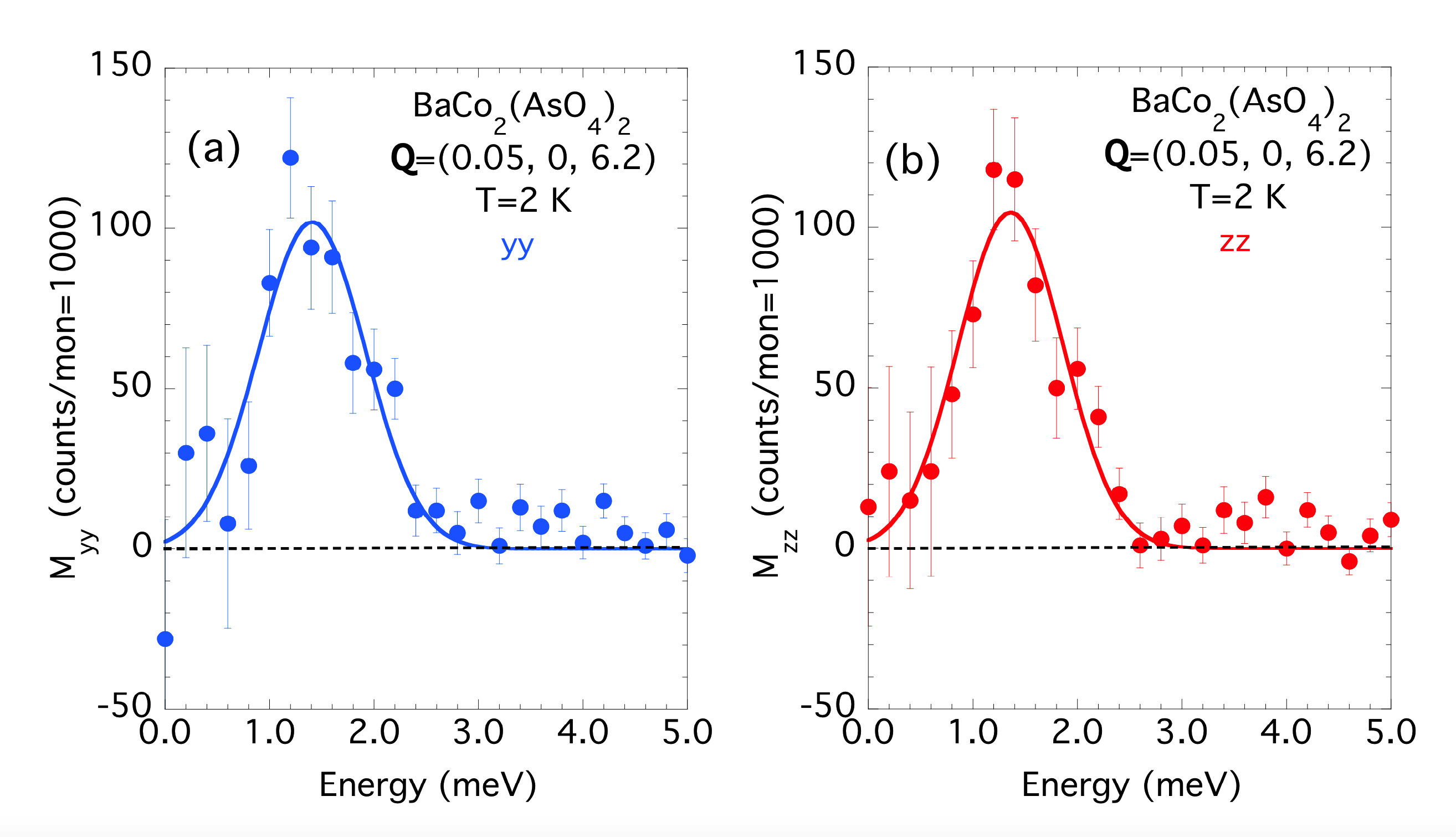}
\caption{Energy-dependence of the pure-magnetic components 
(a): $M_{yy}$ and (b): $M_{zz}$, at the scattering vector $\textbf{Q}=(0.05, 0, 6.2)$, showing the isotropic character of magnetic
excitations at  $q_{a} \approx 0$. The solid lines are fit to a Gaussian function, as described in the text.}
\label{Figure-MyMz0.05}
\end{figure}

\begin{figure}
\centering
\includegraphics[width=10cm]{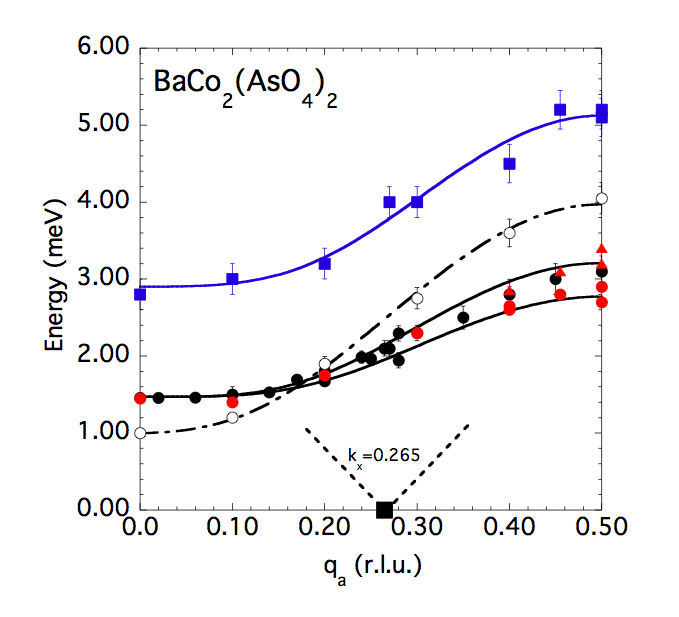}
\caption{Dispersion of magnetic excitations along
${\bf{a}}^{*}$ in zero field (closed symbols) and under a magnetic field of $0.7$ T applied along the b-direction (open circles). The red, orange and blue closed symbols correspond to new data. The black closed symbols correspond to data taken from Ref.~\cite{BCAO-deJongh}. The solid and dashed lines are fit to Eq.~(\ref{BCAO-DispersionCurve}), as described in the text.}
\label{BCAO-SW-dispersion}
\end{figure}

In a first step, we have characterized the coherence of the magnetic ordering developing in BCAO. The wave-vector dependences of magnetic contributions $M_{yy}$ and $M_{zz}$ for several satellite reflections were determined separately from the measurements of the spin-flip (SF) and non-spin-flip (NSF) longitudinal cross-sections $\sigma_{xx}$, $\sigma_{yy}$, and $\sigma_{zz}$, for incident and final neutron polarization selected respectively along the three cardinal directions {\bf x}, {\bf y} and {\bf z}. As it is well known from standard longitudinal polarization analysis (LPA)~\cite{LPA-textbook}, in the absence of chiral and NMI terms, one has $M_{yy}=|\sigma_{xx}-\sigma_{yy}|$ and $M_{zz}=|\sigma_{xx}-\sigma_{zz}|$. Following the notation adopted in section~\ref{SNP}, we define as $M_{a^*}$, $M_{b}$ and $M_{c}$ the elastic structure factors associated with moment components along the {\bf{a}}$^*$, {\bf{b}} and {\bf{c}} directions ($M_{{a^*}{a^*}}=M_{a^*}(M_{a^*})^*$, $M_{{b}{b}}=M_{b}(M_{b})^*$ and $M_{{c}{c}}=M_{c}(M_{c})^*$), respectively. For a given scattering vector ${\bf{Q}}$, one has $M_{y}=M_{a^{*}}\sin(\alpha)+M_{c}\cos(\alpha)$ (where $\alpha=({\bf{Q}}, {\bf{a}^*})$) and $M_{z}=M_{b}$, relations from which one can easily derive the $M_{a^{*}a^{*}}$, $M_{bb}$ and $M_{cc}$ structure factors. As an example, Fig.~\ref{Figure-ScanElastic}(a) shows the dependence on the wave-vector component along ${\bf{a}}^{*}$ of the elastic magnetic contribution $M_{bb}$ for satellite $(0, 0, 6)^{+} \approx (0.27, 0, 4.67)$, while Fig.~\ref{Figure-ScanElastic}(b) shows a similar scan across the scattering vector ${\bf{Q}}=(0.27, 0, 6.1)$, located close to the middle point of the $(0, 0, 6)^{+}$ and $(0, 0, 9)^{+}$ magnetic satellite reflections. For the former, a narrow peak of width (FWHM) $\Delta{q_{a}} \approx 0.0175$ r.l.u., larger than the instrumental resolution, is observed. Assuming for $M_{bb}(q_{a})$ a simple Lorentzian function of half-width $\Gamma_{a^*}=1/\xi_{a*}$, after instrument-resolution correction (${\Delta{q_a}}^{res} \approx 0.013$ r.l.u. (FWHM), as determined from the $(0, 0, 3)$ and $(0, 0, 6)$ structural Bragg reflections), one obtains a coherence length $\xi_{a^{*}} \approx 150$ \AA $\approx 30a$, which indeed is rather short. On Fig.~\ref{Figure-ScanElastic} (b), we show a $Q$-scan along ${\bf{a}}^*$ performed across the  scattering vector ${\bf{Q}}=(0.27, 0, 6.1)$ (located close to the magnetic Brilloun-zone boundary), at  a temperature of $1.5$ K. A broad, purely magnetic contribution representing about $8-9$\% of the contribution ${\bf{Q}}=(0.27, 0, 4.67)$ is clearly observed, with a width (FWHM) $\Delta{q_a} \approx 0.084$ r.l.u., which is much larger than that at ${\bf{Q}}=(0.27, 0, 4.67)$, this being explained as an effect of the finite coherence length along ${\bf{c}}$. A precise value of the interlayer magnetic coherence length $\xi_{c}$ has been obtained from a xx-SF $Q_{c}$-scan across the $(0, 0, 6)^{+}$ magnetic satellite reflection (see Fig.~\ref{Figure-Scan-Qc} (a)), probing the total magnetic contribution. From the width (FWHM) $\Delta q_{c} \approx 0.17$ r.l.u. (again found larger than the resolution width (FWHM) ${\Delta q_{c}}^{res} \approx 0.12$ r.l.u., as measured on the  $(0, 0, 3)$ and $(0, 0, 6)$ structural reflections), again assuming a lorentzian scattering function of $q_{c}=Q_{c}-(6+k_{z})$ of half-width $\Gamma_{c}=1/\xi_{c}$, one has deduced a very short coherence length $\xi_{c}=\frac{c}{\pi \Delta{q_c}} \approx 3c$ , representing about 9-10 correlated layers, only. Figure~\ref{Figure-Scan-Qc} (b) shows a zoom of unpolarized elastic $Q_{c}$-scans across the $(0, 0, 9)^{+}$ satellite reflection performed at several temperature located from both sides of the transition temperature, which confirm the presence of a magnetic contribution over the full Brillouin zone along ${\bf{c}}$ at low temperature, and the quasi-2D behavior of the magnetism above $T_{N}$. \\
The $Q$-dependences of the $M_{bb}$ contribution can be understood by assuming a scattering function modeled by an anisotropic Lorentzian function centered at the satellite-reflection position, $M_{bb}(q_{a} ,q_{c}) \propto f^{2}(Q)/[1+(\frac{q_{a}}{\Gamma_{a^*}})^2+(\frac{q_{c}}{\Gamma_{c}})^2]$, where $q_{a} = Q_{a}-k_{x}$, $q_{c}=Q_{c}-(6+k_{z})$, and $f(Q)$ is the magnetic form factor of  the Co$^{2+}$ions. From the relation giving the $q_{c}$,-dependence of the width (FWHM) along ${\bf{a}^*}$, $\Delta q_{a^*}\approx 2\Gamma_{a^{*}}\sqrt{1+(\frac{q_{c}}{\Gamma_{c}})^2} $, one determines  for $q_{c} \approx 1.43$ r.l.u., $\Delta q_{a} \approx 0.13$ r.l.u., a value which is only in qualitative agreement with the experimental determination. Though in principle not essential for such measurements, at least the use of the polarized neutron diffraction was crucial to prove unambiguously the magnetic nature of contributions, and separate the various magnetic components. As previously reported in Refs.~\cite{BCAO-deJongh} and \cite{BCAO-LPR83}, the ordering along the $b$-direction appears much better established. The Bragg peaks display resolution-limited FWHM, implying a magnetic coherence length along the pseudo chains $\xi_{b}>400$ {\AA} ($\frac{\xi_{b}} {b}>80 $). Obviously, in BCAO a true long-range magnetic ordering is lacking far below $T_{N}$, both along the $c$- and $a^*$-axis. The limited coherence lengths $\xi_{c}$ and $\xi_{a^*}$ can be accounted for by the existence of numerous stacking-faults and/or low-energy  defects, resulting mainly from the strongly frustrated and quasi-2D character of magnetic interactions in this compound. This tendency is even more pronounced in the ferrimagnetic phase between $H_{c1}$ and $H_{c2}$, for which the coherence length along ${\bf{c}}$ decreases down to one interlayer distance ($\frac{\xi_{c}}{c} \approx 1$), while the coherence length along  ${\bf{a}}^*$ is only slightly reduced~\cite{BCAO-deJongh,BCAO-HT}.\\  
We have checked carefully the existence of higher-order harmonics of the modulation, by performing scans along {\bf{a}}$^*$ and {\bf{c}}$^*$. Figure~\ref{Figure-Harmonics} shows two $Q_{a}$-scans across the scattering vectors ${\bf{Q}}=(0.19, 0, 4.9)$ (harmonic $3{\bf{k}}_{1}$ ) and ${\bf{Q}}=(0.36, 0, 4.2)$ (harmonic $5{\bf{k}}_{1} $), which unambiguously show the existence of small and structureless contributions, representing respectively about $0.7\%$ and $0.6\%$ of the main satellite intensity, more than one order of magnitude smaller than those expected for a perfect squared-up modulation (awaited at $11\%$ and $4\%$, respectively). Unpolarized $Q_c$-scans across the scattering vectors ${\bf{Q}}=(0.19, 0, 1.9)$ (corresponding to the third harmonic) from both sides of $T_{N}$, show also structureless, essentially flat magnetic signals after correction by the Co$^{2+}$ form factor and the geometrical factors (see Figure~\ref{Figure-Harmonics-Qc}). Clearly, in BCAO there are no long-range ordered third and fifth harmonics. Indeed, the existence of such a disorder has an important consequence for the determination of the ground-state magnetic structure: Due to the strong broadening of magnetic satellites (especially along the $c$-axis), the standard integrated-intensity method is difficult to apply, this justifying the use of more sophisticated and accurate techniques, like, e.g., the SNP, for the determination of various magnetic structure factors.\\
In Table~\ref{Table-1} we give the polarization matrices for the $(0, 0, 3)$, $(0, 0, 9)$ and $(1, 0, 1)$ pure structural Bragg reflections. For such contributions, one expects no polarization change after scattering and thus a pure diagonal polarization matrix
with equal matrix elements $P_{xx} \approx P_{yy} \approx P_{zz} \approx P_0$. In the real case, the small off-diagonal terms provide us directly with estimates of the accuracy of CRYOPAD (in most of cases smaller than $\pm 0.025$).  
Typical polarization matrices for the magnetic satellite reflections $(0, 0, 0)^{+} \approx (0.265, 0, -1.33)$, $(0, 0, 6)^{+} \approx (0.265, 0, 4.67)$, $(0, 0, 9)^{+} \approx (0.265, 0, 7.67)$ and $(1, 0, 1)^{-} \approx (0.735, 0, 2.33)$ are shown in Table~\ref{Table-2}. Several interesting, model-independent informations can be obtained from the qualitative analysis of diagonal and off-diagonal terms. First, the fact that for any satellite reflections $P_{xx} \approx -P_{0}$ and $P_{yx} \approx P_{zx} \approx 0$,  implies, after Eqs.~(\ref{Eq-Pxx-Mag}) and (\ref{Eq-Pyx-zx-Mag}), that the corresponding elastic chiral terms must be very small (typically $\frac{M_{ch}}{\sigma_{M}} \ll 0.02$), indeed a result still consistent with the simple planar helix structure with two equivalent domains of opposite (${\bf{k/{-\bf{k}}}}$) helicities, as previously determined (see Ref.~\cite{BCAO-deJongh}) and shown in Fig.~\ref{Figure-HelicalStructure}. Second, from Eq.~\ref{Eq-Pyy-Mag}, the fact to have ${\mid}P_{yy}{\mid} \approx {\mid}P_{zz}{\mid} \approx P_{0}$ both for scattering vectors almost parallel to ${\bf{a}}^{*}$ or ${\bf{c}}$, implies $M_{cc} \ll M_{bb}$ (a result expected from the strong planar character), and also $M_{a^*a^*} \ll M_{bb}$. The latter result rules out the helical structure, since for such a structure one should rather have $M_{a^*a^*} \approx M_{bb}$. Instead, our SNP results can be accounted for by assuming a quasi-collinear structure, with magnetic moments pointing along the {\bf{b}} direction, however slightly tilted in order to explain the finite value of the $M_{a^*a^*}$ component. Third, non-zero matrix elements $P_{yz}$ and $P_{zy}$ (with $P_{yz} \approx P_{zy}$) are unambiguously observed for most of magnetic satellite reflections. We have also established that $P_{yz}({\bf{Q}})$ and $P_{zy}({\bf{Q}})$ are both antisymmetric functions of ${\bf{Q}}$: $P_{yz}({\bf{Q}}) \approx -P_{yz}(-{\bf{Q}})$ (and a similar relation for $P_{zy}$). More quantitatively, for the $(1, 0, 1)^{-}$ satellite reflection one obtained the rather strong value ${\mid}P_{yz}{\mid} \approx {\mid}P_{zy}{\mid} \approx 0.17$, whereas for the $(0, 0, 9)^{+}$ satellite reflection (almost parallel to ${\bf{c}}$),  one has determined ${\mid}P_{yz}{\mid} \approx {\mid}P_{zy}{\mid} \approx 0.06$. The former value, in particular, can only be understood from the existence of a finite OP Fourier-component, associated with a non-negligible canting of magnetic moments out of the basal plane, which question the planar character. 
The $Q_{a}$ and $Q_{c}$  dependence of matrix element $P_{yz}$ is summarized in Fig.~\ref{Figure-Pyz(Qc)}. Following the methodology suggested in section~\ref{SNP}, these data have been obtained by measuring $P_{yz}(\bf{Q})$ and $P_{yz}(-\bf{Q})$ (i.e., after rotation of the sample by 180$^{\circ}$) and taking their anti-symmetric combination. Applying Eqs.~(\ref{Eq-Pxx-Mag})-(\ref{Eq-Pyz-Mag}) (valid for pure magnetic contributions), $P_{yx}$, $P_{zx}$, $P_{yz}$, $P_{zy}$, $P_{yy}$ and $P_{zz}$ (the matrix elements useful for the structure determination) can be rewritten as a function of $M_{a^*a^*}$, $M_{bb}$, $M_{cc}$, $M_{a^{*}b}$ and $M_{cb}$ structure factors:
\small {
\begin{eqnarray}
P_{yz} &=& P_{zy} \approx 2\frac{\Re(M_{a^*b})\sin(\alpha)+\Re(M_{cb})\cos(\alpha)}{M_{ip}+M_{bb}}P_{0}  
\label{Eq-term-yz}\\
P_{yx} &=& P_{zx} \approx 2\frac{\Im(M_{a^*b})\sin(\alpha)+\Im(M_{cb})\cos(\alpha)}{M_{ip}+M_{bb}}
\label{Eq-term-chiral}\\
P_{yy} &=& -P_{zz} \approx \frac{M_{ip}-M_{bb}}{M_{ip}+M_{bb}}P_{0}
\label{Eq-term-yy}
\end{eqnarray}
}
in which $M_{ip}=M_{a^*a^*}\sin^2(\alpha)+M_{cc}\cos^2(\alpha)+2\Re(M_{a^*b})\sin(\alpha)\cos(\alpha)$.
The experimental data listed in Table~\ref{Table-2} and those plotted in Fig.~\ref{Figure-Pyz(Qc)} immediately tell us  that $\Re(M_{a^*b})$ and $\Re(M_{cb})$ should be both finite, whereas $\Im(M_{a^*b})$ and $\Im(M_{cb})$ should vanish, as the chiral terms do. In the following, we will consider magnetic structures in which the magnetic moments are almost equal in amplitude. This hypothesis is justified by several reasons. Firstly, the IP anisotropy is relatively weak and should not be relevant, unlike, e.g., for the ANNNI-model case~\cite{Selke-1980,ANNNI-model,Shirakura-2014}. Secondly, our measurements have been performed below 2K, and at such low temperatures (for entropy reasons) fixed-length magnetic moments are expected. Thirdly, constant-amplitude moments are unambiguously observed for the field-induced collinear ferrimagnetic structure at $H=0.4$ T~\cite{BCAO-deJongh,BCAO-HT}, which is described by a similar IP propagation vector, ${\bf{k}}_{ip}=(1/3, 0)$ (following the sequence ${\dots} {\uparrow} {\uparrow} {\downarrow} {\uparrow} {\uparrow} {\downarrow} {\dots}$), and which should have a magnetic energy very close to that of the ground-state structure. Yet, any squaring-up of the modulation should give rise to odd-harmonics Fourier components, with amplitudes ${\mid}m^{b}_{(2p+1)k}{\mid} \approx \frac{m_{b}}{2p+1}$. For BCAO, we think that the extreme weakness of all higher-order harmonic satellites could result directly from the short-ranged nature of spin arrangements along the ${\bf{a}}^*$ and ${\bf{c}}$ directions.  Indeed, the odd harmonics should have a pronounced quasi-2D or even quasi-1D character, which should contribute to reduce their intensities much below their respective ideal squaring-up values, $I_{(2p+1)k} \propto \frac{m_{b}^2}{(2p+1)^{2}}$. \\
At first, the magnetic structure of BCAO can be inferred by only considering the magnetic intensities associated with the wave vector ${\bf{k}}_{1}$. For this wave vector, the magnetic-moment components along ${\bf{a}}^*$, ${\bf{b}}$ and ${\bf{c}}$ can be described by sine-wave sequences, following the relations (the origin of phases being taken w.r.t. the $b$-axis):
\begin{eqnarray}
m_{{a^*}}(\textbf{R}_{ni})&=&m_{b}\sin(\gamma_{i})\cos(2 \pi \textbf{k}.\textbf{R}_{n}+\phi_{i}^{a})  \label{Eq-ma}  \nonumber     \\
m_{b}(\textbf{R}_{ni})&=&m_{b}\cos(\gamma_{i})\cos(2 \pi \textbf{k}.\textbf{R}_{n}+\phi_{i}^{b})  \label{Eq-mb}  \nonumber    \\
m_{c}(\textbf{R}_{ni})&=&m_{c}\cos(2 \pi \textbf{k}.\textbf{R}_{n}+\phi_{i}^{c})  \label{Eq-mc}  \nonumber
\end{eqnarray}
In these relations (which also describe the helical case, by taking $\phi_{i}^{b}=\phi_{i}^{a}-\frac{\pi}{2}$ and $\gamma_{i}=\frac{\pi}{4}$), ${\bf{k}}$ is the propagation vector (in r.l.u.), indices $n$ and $i$ label respectively the cell number and the Bravais-sublattice number ($i=1,2$ for BCAO), $\phi_{i}^{a,b,c}$ ($i=1,2$ ) are the various phase angles for the $i$th Bravais-sublattices and $\gamma_i$ ($i=1,2$) are two tilts w.r.t. the $b$-axis, which possibly introduce some non-collinearity in the structure. 
Assuming  $\gamma_{1}$ and $\gamma_{2}$ small, one can derive analytical expressions for the various contributions involved in Eqs.~(\ref{Eq-term-yz}) and~(\ref{Eq-term-chiral}) (for the hexagonal unit cell with z=3 chemical formula), for satellite reflections $(h_{N}, 0, l_{N})^{\pm}$ :
\begin{eqnarray}
\Re(M_{a^*b})&\approx& \frac{3m_{b}^2}{2}[(\gamma_{1}+\gamma_{2})\cos(\Psi_{ab})C_{b}+(\gamma_{1}-\gamma_{2})\sin(\Psi_{ab})S_{b}]C_{a} 
\label{Eq-Reab} \\
\Im(M_{a^*b})&\approx& \frac{3m_{b}^2}{2}[(\gamma_{1}+\gamma_{2})\sin(\Psi_{ab})C_{b}-(\gamma_{1}-\gamma_{2})\cos(\Psi_{ab})S_{b}]C_{a}  
\label{Eq-Imab} \\
\Re(M_{cb})&=& 3(n_{+}-n_{-}) m_{b}m_{c}C_{b}C_{c} {\cos}(\Psi_{bc})  
\label{Eq-Rebc} \\
\Im(M_{cb})&=& 3(n_{+}-n_{-}) m_{b}m_{c}C_{b}C_{c}{\sin}(\Psi_{bc}) 
\label{Eq-Imbc} \\
M_{a^*a^*}&\approx& 3m_{b}^2[(\frac{\gamma_{1}-\gamma_{2}}{2})^2+\gamma_{1}\gamma_{2}C_{a}^2]     \\
M_{bb}&=& 3m_b^2C_b^2  
\label{Eq-Mbb} \\
M_{cc}&=& 3m_c^2C_c^2   
\label{Eq-Mcc} \\
 \nonumber
 \end{eqnarray}
where $C_{a,b,c}=\cos(\Phi_{a,b,c})$, $S_{b}=\sin(\Phi_{b})$, with $\Phi_{a,b,c}=\pm \frac{ \phi_{a,b,c}}{2}\mp\frac{2 \pi k_{x}}{3} + \frac{2 \pi h_{N}}{3}$ (following ${\bf{Q}}={\bf{H_{N}}\mp {\bf{k}}}$),
$\phi_{a,b,c}=\phi^{a,b,c}_{2}-\phi^{a,b,c}_{1}$, $\Psi_{bc}=\frac{[ (\phi_{1}^{c}-\phi_{1}^{b})+(\phi_{2}^{c}-\phi_{2}^{b}) ]}{2}$, and $\Psi_{ab}=\frac{[ (\phi_{1}^{a}-\phi_{1}^{b})+(\phi_{2}^{a}-\phi_{2}^{b}) ]}{2}$. In these expressions, n$_{+}$ and n$_{-}$ are the populations of anti-phase domains of opposite canting angles, ${\beta
=\pm \arctan}(\frac{m_{c}}{m_{b}})$. The various terms do not depend explicitly neither on $l_N$ (the dependences on $Q_c$ of various matrix elements are only due to the $\sin(\alpha)$ and $\cos(\alpha)$ geometrical factors), nor on the form factor of Co$^{2+}$ ions ($f(Q)$ is assumed to be isotropic by lack of more precise knowledge). For the sake of completeness, we mention that the antisymmetry in ${\bf{Q}}$ of terms $P_{yz}$ and $P_{zy}$ follows directly from the above equations. The various parameters determining the magnetic structure have been derived from the quantitative analysis of polarization matrices measured on several satellite reflections.\\
First, from Eqs.~(\ref{Eq-Imab}) and (\ref{Eq-Imbc}), the smallness of terms $\Im(M_{a^*b})$ and $\Im(M_{bc})$, irrespective of ${\bf{Q}}$, implies that $\Psi_{ab} \approx 0$, $\Psi_{bc} \approx 0$ and $\gamma_{1} \approx \gamma_{2}$. $\Psi_{ab} \approx 0$ and $\Psi_{bc} \approx 0$ imply the following relationship between the various phase angles for components $a$, $b$ and $c$: $\phi_1^b - \phi_1^c=-(\phi_2^b - \phi_2^c)$, and $\phi_1^a - \phi_1^b=-(\phi_2^a - \phi_2^b)$. In other words, the phase differences $\phi_{i}{^a}-\phi_{i}{^b}$ and $\phi_{i}{^b}-\phi_{i}{^c}$ for the Bravais sublattices $1$ and $2$ are alternating.\\
In a second step, the experimental data for component $P_{yz}$ (shown in Fig.~\ref{Figure-Pyz(Qc)}) were self-consistently analyzed by applying Eq.~(\ref{Eq-term-yz}), combined to Eq.~(\ref{Eq-term-yy}) and Eqs.~(\ref{Eq-Reab})-(\ref{Eq-Mcc}). The best fit is realized with the following parameters: $\phi_{a} = 95 \pm 7^{\circ}$, $\phi_{b} = 83 \pm 5^{\circ}$, $\phi_{c} \approx 135 \pm 10 ^{\circ}$, $\Psi_{bc}  \approx \Psi_{ab} \approx 0$,  $\gamma_{1} \approx \gamma_{2} = 2.5{\pm}0.5^{\circ}$, $\frac{m_{c}}{m_{b}} = 0.10 \pm 0.02$, and non-equivalent domain populations, $n_{+} =90 {\pm} 5$\% and $n_{-}=10 {\pm} 5$\% ($n_{+}-n_{-} \approx 80$\%), taking $P_0 \approx 0.91$ and $2 \pi k_x \approx 97^{\circ}$. Consistently, the present $\phi_{b}$ value is in good agreement with the previous determination from unpolarized neutron measurements, which indeed were mainly probing the $M_{bb}$ components~\cite{BCAO-deJongh}. The OP moment  component amounts to $m_{c}  \approx  0.25 {\mu}_{B}$ (taking for the IP component $m_{b}  \approx  2.5 {\mu}_{B}$, after Ref.~\cite{BCAO-deJongh}) and the OP canting angle is ${\beta=\arctan}(\frac{m_{c}}{m_{b}})=5.7 \pm 0.5^{\circ}$. The various lines in Fig.~\ref{Figure-Pyz(Qc)} have been calculated from the above parameters. The agreement between the calculated curves and the experimental data looks reasonably good. However, the fine analysis reveals that the $Q_{a}$-dependence of $M_{cb}$ is better accounted than that of $M_{a^*b}$. Although the magnetic defects at the origin of the very limited coherence length $\xi_{a}$ may explain this difference, it could also mean that the spin arrangement along the $a^*$-direction is more complicated.
Finally, we correlate the well-marked asymmetry between $n_{+}$ and $n_{-}$ to the fact that our sample was an as-grown almost untwined (${\bf{c}}$/-${\bf{c}}$) single crystal.\\
The arrangement for the magnetic-moment component along ${\bf{b}}$ with odd-harmonics up to the 9th order is shown in Fig.~\ref{BCAO-modulation}. As is can be seen, in this non-ideal case more or less regular defects are present (associated with a phase shift of roughly $\pm 2\pi k_{x}$, in order to compensate the anomalously small value of the magnetic moment), with an inter-defect distance $d_{a}  \approx  13a \approx \frac{1/4}{(k_x-1/4)}a$. In BCAO, this distance very likely fluctuate and the phase of the modulation is finally lost over distances corresponding to the correlation length $\xi_{a} \sim 4d_{a}$. This might also explain the lack of long-range order of higher-order (${3\bf{k}}$, ${5\bf{k}}$, ...) harmonics. Finally, the presence of such more or less regular defects explains the existence of an incommensurate propagation vector.  Thus, the in-plane magnetic structure of BCAO can be described as a stacking of quasi-ferromagnetic chains running along to the  ${\bf{b}}$ axis, following the sequence ${\dots} \uparrow \uparrow \downarrow \downarrow \uparrow \uparrow {\dots}$  over a finite length scale, involving magnetic moments of almost constant amplitude, roughly parallel to ${\bf{b}}$ and slightly canted away from the (${\bf{a}}$, ${\bf{b}}$). The ordering of the OP component, driven by the IP one, follows a quite similar sequence, giving rise to the idealized magnetic structure reported in Fig.~\ref{Figure-StructMagnCollinear}. We will come back to this result latter in the discussion.\\
The SNP formalism, summarized by the general Blume-Maleyev equations~\cite{Maleyev-SNP-63,Blume-SNP-63}, can be applied to the analysis of the inelastic contributions, as well. More precisely, the systematic measurements of diagonal and off-diagonal elements of the polarization matrix (especially $P_{yx}$, $P_{zx}$, $P_{yz}$, and $P_{zy}$, see Eqs.~(\ref{Eq-Pxx-Mag})-(\ref{Eq-Pyz-Mag})), may bring interesting new pieces of information, since in principle they allow the determination of the whole cross-sections. In BCAO, one of the key points is the understanding of the nature of the magnetic excitations, especially their relationship with the incommensurate magnetic structure. More specifically, one question which should be addressed is to determine whether the magnetic excitations are simple spin waves or new, more exotic (e.g., multi-particle bound-state or roton-like) excitations. The dispersion of magnetic excitations in BCAO has been first determined from both unpolarized and polarized inelastic neutron scattering experiments. Typical constant-Q scans obtained within the PG-PG (unpolarized) monochromator-analyzer configuration at fixed $k_{f}=1.97$ \AA$^{-1}$ are shown in Fig.~\ref{BCAO-MagnonGroup}(a) and~\ref{BCAO-MagnonGroup}(b), for scattering vectors ${\bf{Q}}=(q_{a}, 0, 6.1)$, with $q_{a}$ spanning the [0, 0.5] half Brillouin zone. The scan at ${\bf{Q}}=(0, 0, 6.1)$ (see Fig.~\ref{BCAO-MagnonGroup}(a)) shows the sharp-gap feature at $\Delta_{0} \approx 1.47$ meV, and another contribution having a maximum intensity at about $3$ meV $\approx 2\Delta_{0}$ and extending up to $6$ meV, characteristic of a 2-particle continuum. The lower-energy contribution, degenerated at small $q_{a}$, splits into two distinct modes at $q_{a}\gtrsim 0.35$. The maximum splitting $ (\approx 0.4$ meV) is observed at the Brillouin-zone boundary ($q_{a}=0.5$), as documented by the energy-scan at  ${\bf{Q}}=(0, 0, 6.1)$ (Fig.~\ref{BCAO-MagnonGroup}(b)). For the latter scattering vector, a third mode peaked at an energy of $5.2$ meV and extending up to about $7$ meV is observed, attributed to the dispersion of the 2-magnon continuum. In agreement with previous measurements, the structure factor of magnetic excitations is maximum at $q_{a} \approx 0$ (and not at $q_{a} \approx k_{1x}$), and decreases rapidly for $q_{a} \gtrsim 0.35$. Although the magnetic origin of all these contributions might follow from their disappearance at high temperature, as shown from the scan at ${\bf{Q}}=(0.5, 0, 6.1)$ performed at $T=100 K$ (see Fig.~\ref{BCAO-MagnonGroup}(b)), it has been unambiguously established from polarized neutron inelastic scattering measurements. Constant-${\bf{Q}}$ scans performed with the incident and final neutron polarization successively parallel to ${\bf{x}}$, ${\bf{y}}$ and ${\bf{z}}$ (LPA configuration) have allowed a precise determination of pure magnetic dynamical structure factors  $M_{yy}$ and $M_{zz}$. Typical results are shown in Fig.~\ref{Figure-MyMz0.27} for the scattering vector {\bf{Q}}=$(0.27, 0, 6.1)$. For this position (almost parallel to ${\bf{c}}^*$), the energy dependences of magnetic fluctuations parallel to ${\bf{a}}^*$, ($M_{yy}$, Fig.~\ref{Figure-MyMz0.27}(a)), and parallel to ${\bf{b}}$ ($M_{zz}$, Fig.~\ref{Figure-MyMz0.27}(b)) can be determined separately. A sharp (resolution-limited) excitation peaked at 2.2 meV is clearly observed in both channels. Fig.~\ref{Figure-MyMz0.05} shows similar data obtained at the scattering vector ${\bf{Q}}=(0.05, 0, 6.2)$, located close to the minimum energy of the dispersion curve. In agreement with the previous unpolarized measurements, a resolution-limited ( ${\Delta}E_{FWHM} \approx 1.1$ meV), almost-isotropic excitation is observed around the energy $\Delta_{o} \approx 1.5$ meV. In addition, the scans displayed in Fig.~\ref{Figure-MyMz0.05} show that there is no trace of a ferromagnetic quasi-elastic contribution at ${\bf{Q}}=(0.05, 0, 6.2)$. Finally, the scans depicted in Figs.~\ref{Figure-MyMz0.27}(a) and ~\ref{Figure-MyMz0.27}(b) establish the magnetic nature of the continuum extending up to $6$ meV.\\
Figure~\ref{BCAO-SW-dispersion} summarizes the dispersion of magnetic excitations along the [1 0 0] direction. For the sake of comparison, we have also included data in an applied magnetic field of 7 kG, taken from Ref.~\cite{BCAO-Excitations}. As previously observed, the minimum energy of the excitation spectrum is located at ${\bf{q}}={\bf{0}}$ and not at ${\bf{q}}={\bf{k}}_{1}$. Unexpectedly, the dispersion curves of magnetic excitations in BCAO reflect the proximity to a ferromagnetic ground state, despite the absence of strong ferromagnetic Bragg scattering along the $(0, 0, Q_{c})$ reciprocal-lattice line \cite{BCAO-deJongh,BCAO-Excitations}. The dispersion curves along the [1 0 0] direction have been analyzed from the following empirical relation: 
\begin{equation}
E(q_{a})=\sqrt{\Delta_{0}^{2}+\Delta_{1}^{2}\sin^{2}(\pi q_a)+\Delta_{2}^{2}\sin^{4}(\pi q_a)}
\label{BCAO-DispersionCurve}
\end{equation}
In zero field, the best fit of experimental data to Eq.~(\ref{BCAO-DispersionCurve}) (solid line in Fig.~\ref{BCAO-SW-dispersion}) is achieved with the parameters $\Delta_{0} \approx 1.47$ meV, $\Delta_{1} \approx 0.6$ meV, $\Delta_{2} \approx 2.30$ meV for the lower-energy mode and $\Delta_{0} \approx 1.47$ meV, $\Delta_{1} \approx 0.6$ meV, $\Delta_{2} \approx 2.75$ meV for the upper-energy mode. The dispersion of the maximum energy of the 2-magnon continuum can also be reproduced by Eq.~(\ref{BCAO-DispersionCurve}), with parameters $\Delta_{0} \approx 2.9$ meV, $\Delta_{1} \approx 0.8$ meV and $\Delta_{2} \approx 4.1$ meV. The non-conventional behavior  of the dispersion of the lower-energy modes along [1 0 0]  is well documented by the smallness of  the $\Delta_1$ parameters, this reflecting the quasi absence of a quadratic term (weak stiffness), and the flatness of the dispersion curve for $q_a\lesssim0.15$ r.l.u., indeed very reminiscent of a quasi-1D excitation rather than a quasi-2D one. The situation turns out to be very different in magnetic fields $H>H_{c2}$. As it can be seen in Fig.~\ref{BCAO-SW-dispersion}, in the saturated paramagnetic phase ($H=0.7$ T in the present case), the best fit of data to Eq.~(\ref{BCAO-DispersionCurve}) is obtained with the set of parameters $\Delta_{0} \approx 1.0$ meV (smaller gap energy), $\Delta_{1} \approx 1.7$ meV and $\Delta_{2} \approx 3.4$ meV, which show that the quadratic term, and consequently the propagative character, are recovered.\\ 
As mentioned in section~\ref{intro}, the linear spin-wave theory applied to the $J_{1}-J_{2}-J_{3}$ XXZ (planar) model described by Eq.~(\ref{Eq-Hamiltonian}) is unable to explain the main features of the dispersion curves (especially the gap at $q_{a}=0$), both for the collinear IC ($H<H_{c1}$) and the ferrimagnetic ($H_{c1<}H<H_{c2}$) structures. Instead, above $H_{c2}$ (in the saturated paramagnetic phase described by the wave vector ${\bf{k}} = {\bf{0}}$), the dispersion curves (including the field-dependence of the gap energy at $q_{a} \approx 0$, $\Delta_0(H) \approx S\sqrt{6(J_{1}+2J_{2}+J_{3})}\sqrt{g_{x}\mu_{B}H}$ ) can be quantitatively reproduced by the simple SW theory for the $J_{1}-J_{2}-J_{3}$ XXZ model.\\
In order to quantify the $q_{a}$-dependence of spin-dynamics in BCAO (especially the anisotropy in spin-space of magnetic fluctuations), we have undertaken the determination of full polarization matrices for several inelastic positions covering the first Brillouin zone.  As examples, we give in Table~\ref{Table-3} the polarization matrices determined on the inelastic magnetic contributions at ${\bf{Q}}=(0, 0, 4.67)$ and $\Delta E=1.5$ meV (corresponding to the minimum of the dispersion curve), ${\bf{Q}}=(0.27, 0, 4.67)$ and $\Delta E=2.1$ meV (mostly parallel to {\bf{c}}$^*$, corresponding to ${\bf{q}} \approx {\bf{k}}_{1}$), and finally {\bf{Q}}=$(0.73, 0, 0.8)$ and $\Delta E=$2.3 meV (mostly parallel to {\bf{a}}$^*$). The matrix at ${\bf{Q}}=(0, 0, 4.67)$ and ${\Delta}E=1.5$ meV has a very simple form: Only  $P_{xx}$ is not zero (with $P_{xx} \approx -0.90$), all the other terms being very small, especially $P_{yx}$ and $P_{yz}$. However, it is worth noting that $P_{yy}$ is not equal to -$P_{zz}$, as it should be for a pure magnetic contribution (see Eq.~(\ref{Eq-Pyy-Mag})). In order to account for this fact, we have to assume the existence of a small structural contribution, $N$, superimposed to the magnetic ones. By applying Eqs.~(\ref{Eq-Pxx-N&M})-(\ref{Eq-Pzz-N&M}), the $N$, $M_{yy}$ ($\approx M_{a^*a^*}$) and $M_{zz}$ ($=M_{bb}$)  structure factors can be determined from the self-consistent analysis of the $P_{xx}$,  $P_{yy}$ and $P_{zz}$ terms. Taking  $P_{0} \approx 0.93$, one obtained for the additional structural component a ratio $\frac{N}{M_{bb}}=0.025 \pm 0.014$ , which indeed is at the limit of the experimental accuracy, and a very small anisotropy ratio $\frac{M_{yy}-M_{zz}}{M_{bb}} \approx -0.01\pm 0.010$, which implies a ratio $\frac{M_{{a^*}{a^*}}}{M_{bb}} \approx 0.99$. The magnetic fluctuations along ${\bf{a}}^*$ and ${\bf{b}}$ are quasi isotropic at ${\bf{Q}}=(0, 0, 4.67)$ and ${\Delta}E=1.5$ meV, a result which is at first surprising, owing to the axial character of the ground-state magnetic structure (we have found $\frac{M_{{a^*}{a^*}}}{M_{bb}} \approx 0.02$ for the magnetic satellite reflection $(0, 0, 6)^+$). We will come back to this point later. At ${\bf{Q}}=(0.27, 0, 4.67)$ and $\Delta E=2.1$ meV, the polarization matrix is mainly diagonal, within the error bars. Matrix elements  $P_{yy}$ and $P_{zz}$ now present finite and opposite values, $P_{zz}\approx -P_{yy} = 0.157 \pm 0.015$. Contrary to the matrices determined on the magnetic satellite reflections (see Table~\ref{Table-2}), the $P_{yz}$ and $P_{zy}$  elements are vanishing small, as $P_{yx}$ and $P_{zx}$ are (no antisymmetric chiral correlations). From the values of $P_{yy}$ and $P_{zz}$, the ratio $\frac{M_{yy}}{M_{zz}}=0.71 \pm 0.02$ is deduced, and finally the ratio $\frac{M_{{a^*}{a^*}}}{M_{bb}}=0.86 \pm 0.03$ is determined. Similar measurements at a different configuration (essentially a different Heusler analyzer and a slightly different energy transfer of 2.3 meV) have given a slightly different ratio, $\frac{M_{{a^*}{a^*}}}{M_{bb}}=0.82 \pm 0.03$. Thus, the IP magnetic structure factors at $q_{a} \approx 0.27$ are weakly anisotropic, with a mean ratio $\frac{M_{{a^*}{a^*}}}{M_{bb}}=0.84 \pm 0.02$. Same as for the scattering vector ${\bf{Q}}=(0, 0, 4.67)$, magnetic fluctuations exist both along ${\bf{a}}^{*}$ and ${\bf{b}}$. Within the error bars, no extra structural contribution is detected at ${\bf{Q}}=(0.27, 0, 4.67)$ and $\Delta E=2.1$ meV. However, such a contribution was again observed at ${\bf{Q}}=(0.27, 0, 3.1)$ and $\Delta E=2.3$ meV, with a quite similar intensity ratio, $\frac{N}{M_{bb}}=0.03 \pm 0.015$. For the investigation at ${\bf{Q}}=(0.73, 0, 0.8)$ and $\Delta E=2.3$ meV, we have used the full SNP methodology, prompted by the underlying structural contribution which might give rise to non-zero NMI terms, through the putative existence of hybrid N-M correlation functions.  First, we have measured the polarization creation along the three cardinal directions ${\bf{x}}$, ${\bf{y}}$ and ${\bf{z}}$, starting from an unpolarized beam (in our case produced by a pyrolithic-graphite monochromator). From these measurements, we have determined polarization components, $P_{0x}=0.008 \pm 0.008$, $P_{0y}=0.009 \pm 0.008$, and  $P_{0z}=-0.002 \pm 0.008$. Within the error bars, no polarization of the scattered beam could be detected. After Eqs.~(\ref{Eq-P0x})-(\ref{Eq-P0z}) in section~\ref{SNP}, this result implies necessarily that $\frac{M_{ch}}{M_{bb}} \approx \frac{R_{y}}{M_{bb}} \approx \frac{R_{z}}{M_{bb}}\approx 0$: The magnetic antisymmetric (chiral) dynamical contributions and the symmetric inelastic NM correlation functions are all vanishing small. The polarization matrix given in Table~\ref{Table-3} (bottom) shows several interesting features. From the self-consistent analysis of the $P_{xx}$, $P_{yy}$ and $P_{zz}$ terms, and by applying Eqs.~(\ref{Eq-Pxx-N&M})-(\ref{Eq-Pzz-N&M}), the ratio between the OP and IP dynamical structure factors is derived, $\frac{M_{cc}}{M_{bb}}=0.025\pm 0.005$ (taking  $P_{0}=0.75 \pm 0.03$), consistent with the well-marked planar character found from the neutron diffraction measurements. More interesting, the analysis of matrix elements $P_{xx}$,  $P_{yy}$ and $P_{zz}$ at {\bf{Q}}=(0.73, 0, 0.8) and ${\Delta}E=2.3$ meV again reveals the presence of a non-negligible nuclear (structural) contribution with an absolute intensity, $N=(0.093 \pm 0.010)M_{bb}$, quite similar to the two previous cases if we remember that $M_{bb}$ is smaller at this scattering vector. Table~\ref{Table-4} lists the $yx$, $zx$, $yz$, and $zy$ off-diagonal matrix elements for the two opposite scattering vectors ${\bf{Q}}=(0.73, 0, 0.8)$ and ${\bf{Q}}=(-0.73, 0, -0.8)$, at the same energy transfer ${\Delta}E=2.3$ meV. As emphasized in section~\ref{SNP}, an accurate value of the anti-symmetric NMI term $I_{z}$ can be deduced by considering the anti-symmetric combination of the $yx$ off-diagonal matrix element at scattering vectors $\pm${\bf{Q}} (see Eqs.~(\ref{Eq-Pyx-NMI})-(\ref{Eq-Pzy-NMI})). From the experimental values, one determines the ratio $\frac{I_{z}}{M_{bb}}=-0.007 \pm 0.008$.  The $I_{y}$ antisymmetric NMI term, invariant in a rotation of the sample by $180^{\circ}$, cannot be determined at a similar accuracy, due to the impossibility to cancel the systematic errors. From the $xz$ components given in table~\ref{Table-4}, one got the average ratio $\frac{I_{y}}{M_{bb}} = -0.035 \pm 0.008$, which looks finite within the error bars, but could originate from the systematic errors introduced by CRYOPAD. At least this value is not much different than the average value ${\mid}\frac{P_{xz} + P_{zx}}{2} {\mid} \approx 0.036$ found from the measurements of the polarization matrix on the magnetic Bragg satellite $(0.73, 0, 2.33)$ (see Table~\ref{Table-2}), expected to be null in the present case. Thus, in BCAO the symmetric ($R_{y}$ and $R_{z}$) and the antisymmetric ($I_{y}$ and $I_{z}$) NMI terms seem all vanishing small at the accuracy of our measurements, this showing the absence of any dynamical cross-correlation function coupling the structural  and magnetic degrees of freedom, which indeed are passively coexisting. We have no clear explanation about the origin of the additional structural contribution detected by SNP at ${\bf{Q}}=(0.73, 0, 0.8)$, which might be due to the existence of irrelevant crystallographic defects in the investigated single crystal. Finally, we worth note that finite off-diagonal matrix elements $P_{yz}$ and $P_{zy}$ are also detected at the scattering vector ${\bf{Q}}=(0.73, 0, 0.8)$, with $P_{yz} \approx P_{zy}$, and roughly antisymmetric in ${\bf{Q}}$ (see Table~\ref{Table-4}). From their antisymmetric combinations, $P_{yz}^{a}$ and $P_{zy}^{a}$, accurate values of the $P_{yz}$ and $P_{zy}$ matrix elements have been deduced, ${\mid}P_{yz}{\mid} \approx {\mid}P_{zy}{\mid} = 0.075{\pm}0.025$. According to section~\ref{SNP}, in order to explain such non-zero off-diagonal matrix elements, one should invoke the existence of a rather-strong correlation function coupling the inelastic OP and IP magnetic fluctuations. In the next section, we will prove more quantitatively that in BCAO such a correlation function is indeed a quite {\it{trivial}} one, which essentially originates from the strong anisotropic character of precessions of Co$^{2+}$ magnetic moments.\

\section{Discussion \& Conclusion}
\label{DiscussionConclusion}

\begin{figure}
\centering
\includegraphics[width=10cm]{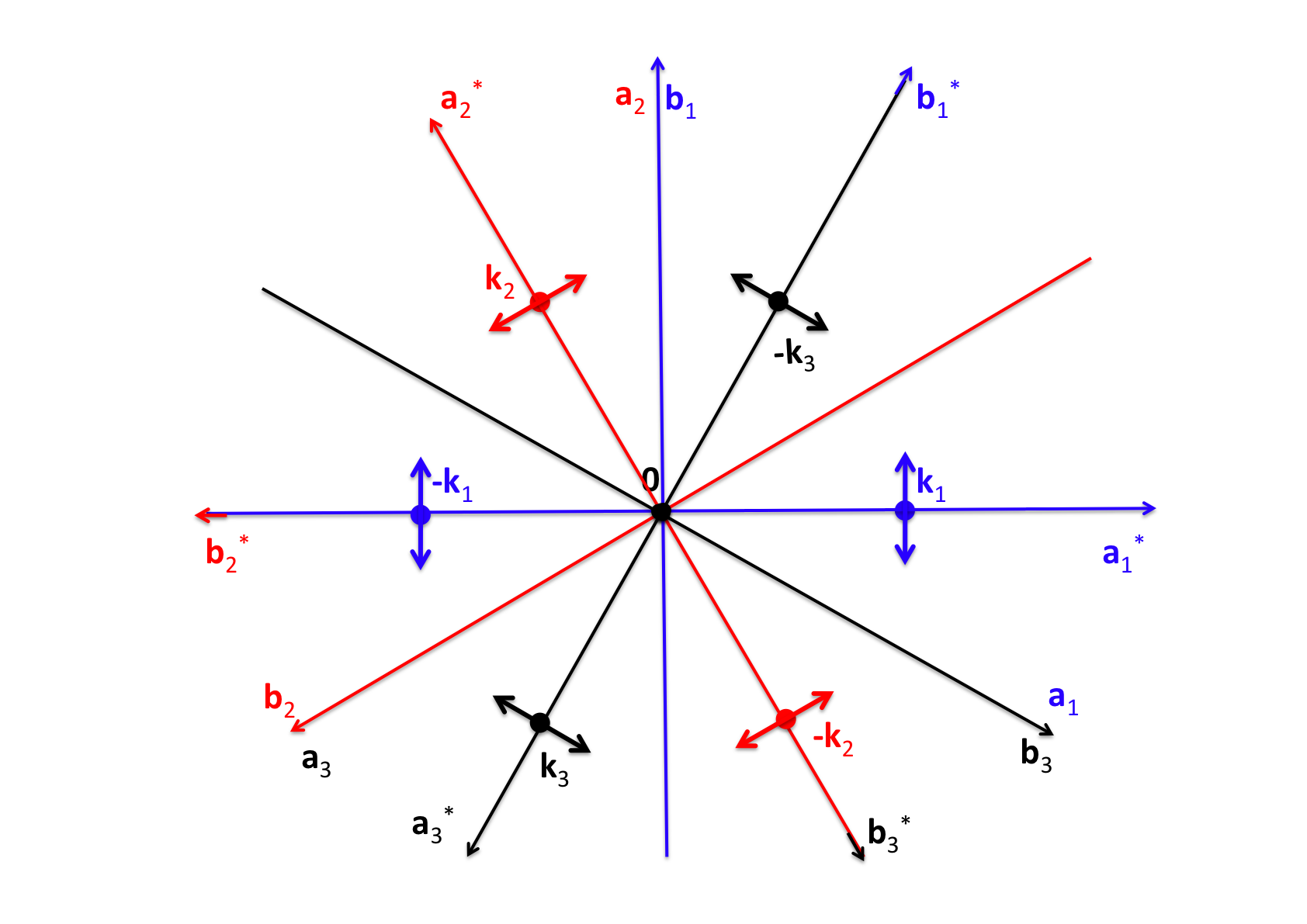}
\caption{Projection along the ${\bf{c}}$-direction of the k-domain structure in \BCAO. The arrows indicate the various moment directions. For spin wave-like excitations, the magnetic fluctuations are quasi-2D and perpendicular (transverse) to the magnetic moments.}
\label{k-domain}
\end{figure}

\begin{figure}
\centering
\includegraphics[width=7cm]{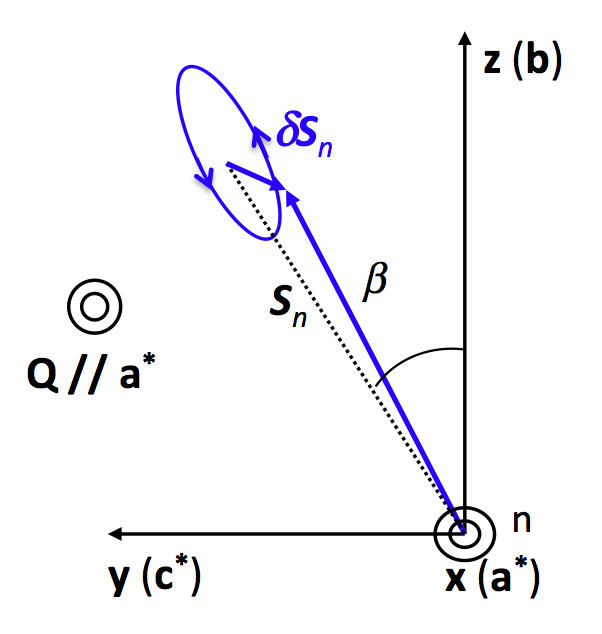}
\caption{Anisotropic moment precessions in {\BCAO} (projection in the (${\bf{b}}$, ${\bf{c}}$) plane).}
\label{Figure-AnisotropicPrecessions}
\end{figure}

Our comprehensive investigation by SNP and LPA of elastic and inelastic contributions in BCAO has brought new and very relevant pieces of information concerning the magnetic ordering and the spin dynamics of this complicated compound. A first unexpected result is the discovery that the magnetic structure of BCAO is collinear, with magnetic moments roughly aligned along the $b$-axis and a non-negligible out-of-plane component, associated with a canting  of about 5.7$^\circ$ w.r.t. the (${\bf{a}}$, ${\bf{b}}$) plane. In BCAO, the magnetic ordering is not very well established both along ${\bf{a}}^*$ ($\xi_{a^*}/a \sim 30$) and ${\bf{c}}$ ($\xi_{c}/c \sim 3$), being at much longer range along ${\bf{b}}$. Indeed, there is a puzzling paradox between the apparent disorder of the incommensurate ground-state structure along the ${\bf{a}}^*$ direction, and the existence of well-defined (resolution-limited) excitations, displaying a dispersion relation rather of ferromagnetic type (energy minimum at $q_{a}=0$), which questions directly the nature of magnetic excitations in BCAO. We have also confirmed the unconventional shape of the dispersion relation along the ${\bf{a}}^*$ direction (absence of quadratic term), and established the quasi-isotropic character of the inelastic magnetic response, which contrasts with the marked axial character of the spin arrangement. Obviously, our SNP results on inelastic magnetic contributions rise-up the question of the relationship between the excitation spectrum and the collinear canted ground-state structure. This is the subject of the following discussion.\\ 
From a general point of view, all the structures involved in BCAO (whether simple helix, AF-collinear, ferrimagnetic or saturated paramagnetic) are quasi-2D, or even quasi-1D structures, exhibiting both a high degree of degeneracy and a strong frustration of exchange interactions. The various structure energies should be very close to each other, as it can be inferred from the small values of the various critical fields. Taking into account the pseudo-chain character inherent to all structures, the classical structure energy has the simple form, $E_{s}=E_{c}+e_{ic}$, where $E_{c}$ represents the energy of an isolated quasi-ferromagnetic chain, and $e_{ic}$ is the inter-chain energy. For the idealized ${\dots} {\uparrow} {\uparrow} {\downarrow} {\downarrow} {\uparrow} {\uparrow} {\downarrow} {\downarrow} {\dots}$ ground-state structure of BCAO, one has: 
$E^{GS}_{c} = -2s^2(J_{1} +J_{2})$, and $e^{GS}_{ic} =0$, which corresponds to the energy of an assembly of isolated ferromagnetic chains, irrespective of $J_{3}$. For the ferrimagnetic (Ferri) and saturated-paramagnetic (Ferro) structures, the intra-chain classical energies are the same: $E^{Ferri}_{c} =E^{Ferro}_{c}= -2s^2(J_{1} +J_{2}) = E^{GS}_{c}$, the inter-chain classical energies being $e^{Ferri}_{ic}=s^2(\frac{J_{1}}{3}+\frac{4J_{2}}{3}+J_{3})$ and $e^{Ferro}_{ic}=-s^2(J_{1}+4J_{2}+3J_{3})$, respectively. The weakness of the critical fields $H_{c1}$ (${\propto}(e^{Ferri}_{ic}-e^{GS}_{ic})$) and $H_{c2}$ (${\propto}(e^{Ferro}_{ic}-e^{Ferri}_{ic})$), implies that $e^{GS}_{ic} \lesssim e^{Ferri}_{ic} \lesssim e^{Ferro}_{ic}$ (condition indeed not satisfied from the above classical energies) and that $J_{1}+4J_{2}+3J_{3} \approx (0.04-0.05)J_{1}$) is small. The latter relation can be quantitatively satisfied by taking the ratios $\frac{J_{2}}{J_{1}} \approx -(0.04-0.05)$ and $\frac{J_{3}}{J_{1}} \approx -(0.27-0.28)$, which are not much different than those determined from the analysis of spin-wave dispersions in the saturated paramagnetic phase. Note, however, that with all these exchange-parameters sets, the classical ground state should be ferromagnetic, in disagreement with the experimental results. Obviously, a more accurate treatment of the problem is required, which should at least include the quantum corrections in the energy calculation of the different structure as a function of $J_{1}$, $J_{2}$ and $J_{3}$. Anyway, the weak effective inter-chain couplings allow the rotation, or the change of the magnetic moment length (associated, e.g.,  with a phase shift) of long chain segments at very low energy cost. In particular, for the ground-state structure one can show that a ferromagnetic pseudo-chain located between ${\uparrow}$ and  ${\downarrow}$ pseudo-chains, following the sequence ${\dots} {\uparrow} {\uparrow} {\downarrow} {\downarrow} {\uparrow} {\uparrow} {\nearrow} {\downarrow} {\uparrow} {\uparrow} {\dots}$, can be entirely rotated by an arbitrary angle without any cost in energy (assuming the IP axial anisotropy term weak). As already mentioned, this feature could explain very well the H-T phase diagram of BCAO. The ease of creating low-energy (quasi static) defects is also at the origin of the step-like temperature dependence of the staggered order parameter, ${\mid}{\bf{m}}_k(T){\mid}$. Without demonstration so far, we believe that the high degeneracy of the ground state could also be at the origin of the strong $T^2$-term in the low-T magnetic specific heat, as it is the case, e.g., for the Kagome lattice~\cite{kagome-specifiheat}.\\
On the theoretical side, the classical phase diagram of the $J_{1}-J_{2}-J_{3}$ model on the honeycomb lattice has been investigated three decades ago~\cite{Rastelli79}.  In a very narrow region of $J_{2}/J_{1}$ and $J_{3}/J_{1}$ values, a helical phase described by an IC wave vector ${\bf{k}}=(k_{x}, 0)$ was predicted as the ground-state structure, very close to the ferromagnetic phase. Exact diagonalizations and linear SW calculations, both for antiferromagnetic and ferromagnetic (indeed the BCAO case) n.n. interactions $J_{1}$ recently performed on the $S=1/2$, $J_{1}-J_{2}-J_{3}$ model on the honeycomb lattice have shed some light on the role of frustration~\cite{J1J2J3-Jussieu}. Among other predictions, it was conjectured that for $J_{1}>0$, frustration (be it due to $J_{2}$ or $J_{3}$) could lead to the disappearance of the ferromagnetic phase at the expense of a spin-liquid phase with short-range IP correlations and the opening of spin-gaps in the excitation spectrum, which seem in qualitative agreement with the experimental results in BCAO. The presence of a frustration-enhanced gapped spin-liquid phase for a quantum-spin (spin-$1/2$) system on the honeycomb lattice has also been predicted by Takano~\cite{Takano06} and more recently by Bishop {\it{et al.}}~\cite{Bishop-XY,Bishop-XXZ,Bishop-3,Bishop-4}. Unfortunately, almost nothing exists for the $J_{1}-J_{2}-J_{3}$ planar model on the honeycomb lattice in applied magnetic field. \\
Regarding the small OP canting ($\beta  \approx  6^{\circ}$) and tilt angles ($\gamma  \approx  2.4^{\circ}$), they clearly highlight the complexity of BCAO. At least, they rule out the simple  model developed long ago for the Co$^{2+}$ ion in octahedral, trigonaly-distorted environment~\cite{cobalt2plus}.  Although being able to explain the strong planar anisotropy, this model is unable to account neither for the canting, nor for the tilt angle.This clearly shows that the solution of this problem demands to go beyond the too simple bilinear XXZ Hamiltonian given by Eq.~(\ref{Eq-Hamiltonian}), and that higher-order terms  (like, e.g., the biquadratic, Dzyaloshinski-Moriya or anisotropic exchange terms) must be taken into account.\\
In BCAO, the excitation spectra are rather unconventional and present some puzzling features. As previously reported, the excitation spectra of the (zero-field) ground-state and (field-induced) intermediate ferrimagnetic structures, indeed very similar (see Ref.~\cite{BCAO-deJongh}), cannot be explained from the simple linear 2D SW theory. In addition, they do not better verify the predictions for an incommensurate modulated phase~\cite{IC-Excitations}. Instead, the quasi-absence of a quadratic term in the dispersion along the $a^*$-axis leads us to assume that the small-$q$ excitations in BCAO are quasi-localized modes displaying a well-defined spin-gap of energy $\Delta_0 \approx 1.45$ meV. The origin of such  a strong spin-gap localized at $q_a \approx 0$ remains intriguing and quantitatively not understood. Although the presence of a small easy-axis anisotropy term (wether on-site or due to anisotropic exchange) favoring an alignment of magnetic moments along the $b$-axis in BCAO is certain, the magnitude of the spin-gap cannot be accounted for by such a term, alone. Considering the mostly ferromagnetic character of the spin-excitation spectrum, one should have: $\Delta_{0} \eqsim 6S(J_{1}+2J_{2}+J_{3}) \sqrt{1-\alpha_{z}}\sqrt{\mid \frac{J^{x}_{1}-J^{y}_{1}}{J_{1}}} \mid$. In order to account for the gap-energy value, one would have to postulate a rather strong IP axial anisotropy of various coupling parameters, ${\mid}\frac{J^{x}_{n}-J^{y}_{n}}{J_{n}}{\mid} \approx 0.08$ ($n$=1,2 and 3). Such a high value is definitively inconsistent with the INS results under field, which gave an upper limit at least one order of magnitude weaker (IP anisotropy field $H_{a}^{ip} \lesssim 0.15$ T). As suggested in Ref.~\cite{J1J2J3-Jussieu} for the strongly frustrated honeycomb lattice, the spin-liquid nature of the ground-state could explain the opening of an energy gap in the spin-excitation spectrum. However, the value which is predicted by the numerical simulations is by far too small. If the excitations in BCAO are really associated with quasi-ferromagnetic pseudo chains running along the b-axis, whose spins are coupled through effective interactions $\tilde{J} \approx J_{1}+J_{2}\approx 40$ K (irrespective of $J_{3}$), one has $\frac{\Delta_0}{\tilde{J}} \approx 0.42$, a value surprisingly very close to the predicted Haldane-gap value ($ \approx 0.41$) for the $S=1$ antiferromagnetic Heisenberg chain~\cite{HaldaneGap}. Although this may be a simple coincidence, this remark could in fact reveal the very unconventional and complex nature of the gaped mode at ${\bf{q}}={\bf{0}}$. \\
A quantitative interpretation of our inelastic SNP results (especially the origin of the finite $P_{yz}$ and $P_{zy}$ matrix elements) may emerge from the following simple model. In the very realistic case of a magnetic structure with magnetic moments mainly aligned along the $b$-axis ($\gamma \approx 0$) and canted out of the (${\bf{a}}$, ${\bf{b}}$) plane by an angle $\beta={\pm}{\arctan}(\frac{m_{c}}{m_{b}})$ (see Fig.~\ref{Figure-AnisotropicPrecessions}) undergoing strongly-anisotropic precessions of pulsation $\omega_{0}$ around their average positions, one can calculate the $P_{yx}$ and $P_{yz}$ components. For a given scattering vector ${\bf{Q}}$, neglecting at first the effect of $k$-domains, one can show that the time-dependence of various spin components for transverse fluctuations are given by the following equations:
  \begin{eqnarray}
  S_{a^*}&=&\delta S_{a^*}\cos(\omega_{0} t), \nonumber  \\
  S_b&=&\delta S_c \sin(\beta) \sin(\omega_{0} t), \nonumber \\
  S_c&=&\delta S_c \cos(\beta) \sin(\omega_{0} t), \nonumber
 \end{eqnarray}
The various dynamical structure factor components are calculated by Fourier transform in time of the associated spin components:
\begin{equation}
    M_{ij}(\omega) = \int \limits_{-\infty}^{+\infty} \,S_{i}(t)S^{*}_{j}(0) \, e^{i \omega t} \, dt  \nonumber
\end{equation}
from which one can derive, by applying Eqs.~\ref{Eq-Pyx-zx-Mag} and \ref{Eq-Pyz-Mag}, expressions for the pure magnetic off-diagonal components $P_{yx}$ and $P_{yz}$:
\begin{eqnarray}
  P_{yx} &=& -\frac{2(p_{+}-p_{-}) (n_{+}-n_{-})\sin(\alpha) \sin(\beta)\delta S_{a^*}\delta S_{c}}{\sin^2(\alpha)(\delta S_{a^*})^2+[\sin^2(\beta)+\cos^2(\alpha)\cos^2(\beta)](\delta S_{c})^2}, \label{Pyx-inelastic-a*}  \nonumber \\
  P_{yz} &=&  -\frac{(n_{+}-n_{-}) \cos(\alpha)\sin(2\beta)(\delta S_{c})^2}{\sin^2(\alpha)(\delta S_{a^*})^2+[\sin^2(\beta)+\cos^2(\alpha)\cos^2(\beta)](\delta S_{c})^2}P_{0} \nonumber \label{Pyz-inelastic-Q}
\end{eqnarray}
in which $\delta S_{a^*}$ and  $\delta S_{c}$ represent respectively the (anisotropic) transverse components along {\textbf{a}$^*$} and {\textbf{c}} of precessing moments, related to the OP and IP dynamic structure factors by the relation, $\frac{\delta S_{c}}{\delta S_{a^*}} \sim \sqrt{\frac{M_{cc}}{M_{a^*a^*}}}\approx 0.17$, $p_{+}$ and  $p_{-}$ being respectively
the proportions of clockwise and counterclockwise precessions. The weakness of the $P_{yx}$ components (experimentally $P_{yx} \approx 0$, irrespective of ${\bf{Q}}$, see Table~\ref{Table-3}) can be explained by several factors: $p_{+}  \approx  p_{-}$ (symmetry clockwise/anticlockwise), $\frac{\delta S_{c}}{\delta S_{a^*}} \ll 1$ (planar character of magnetic fluctuations) and the small canting ($\sin(\beta) \ll 1$).\\
For ${\bf{Q}}$ almost parallel to {\bf{a}}$^*$ ($\alpha$ small), our simple model predicts $P_{yx} \approx 0$ (no chiral term $M_{ch})$, as experimentally observed, and a finite $P_{yz}$ term, directly related to the canting angle $\beta$ and the anti-phase domain populations:
\begin{equation}
  P_{yz} \approx -\frac{(n_{+}-n_{-})(\frac{\delta S_{c}}{\delta S_{a^*}})^2}{\sin^2(\alpha)+(\frac{\delta S_{c}}{\delta S_{a^*}})^2} \sin(2\beta) P_{0} \label{Pyz-inelastic-a*}
\end{equation}
Taking $\sin^2(\alpha) \approx 0.04$, $\beta \approx 6^{\circ}$, $P_{0} \approx 0.75$ and $n_{+} - n_{-} \approx 0.8$ as parameters in Eq.~(\ref{Pyz-inelastic-a*}), one obtains ${\mid}P_{yz}{\mid} \approx  0.07$, a value which is in quantitative agreement with the experimental determination, ${\mid}P_{yz}{\mid} \approx  0.08$. For the helical structure depicted in Fig.~\ref{Figure-HelicalStructure}, assuming equally-populated ({\bf{k}}/-{\bf{k}} ) helicity domains, it is easy to show that $P_{yx}=0$ and $P_{yz}=0$. The observation of finite inelastic $P_{yz}$ and $P_{zy}$ components is very important, since it rules out the helical structure as the ground-state structure of BCAO.\\
For {\bf{Q}} almost parallel to ${\bf{c}}^*$ (case $\alpha \approx \frac{\pi}{2}$), one has:
 \begin{equation}
   P_{yz} \approx -(n_{+}-n_{-})(\frac{\delta S_{c}}{\delta S_{a^*}})^2 {\cos(\alpha)} \sin(2\beta) P_{0} \nonumber \label{Pyz-inelastic-c}
\end{equation}
In this case,  the weakness of the observed $P_{yz}$ terms (see table~\ref{Table-3}) can be explained by the conjunction of three factors: $(\frac{\delta S_{c}}{\delta S_{a^*}})^2 \ll 1$ (planar character of fluctuations), $\sin(\beta) \ll 1$ (small canting of magnetic moments) and $\cos(\alpha) \ll 1$. All results together, our inelastic SNP measurements on BCAO are quantitatively accounted by considering the excitations close to $q_{a}=0$ as simple precessions, leading us to conclude that they are conventional spin-waves.\\
One puzzling feature of the magnetic excitation spectrum at $q_{a} \approx 0$ in BCAO concerns the quasi-isotropy of magnetic fluctuations, which is barely understandable from a single $k$-domain axial-type structure as that depicted in Fig.~\ref{Figure-StructMagnCollinear}, unless strong longitudinal fluctuations exist, oddly peaked at the same energy than the transverse ones. Such fluctuations (which could easily originate, e.g., from fluctuations of the various phase angles) are generally associated with two-magnon excitations, and should rather contribute to the magnetic continuum that has been observed between 3 and 5 meV.  Alternately, the very weak anisotropy of dynamical structure factors $M_{a^*a^*}$ and $M_{bb}$ for $q_{a} \lesssim 0.15$ r.l.u. can be explained by taking into account the $k$-domain structure, sketched in Fig.~\ref{k-domain}. For a given wave vector ${\bf{q}}$, the magnetic response will be the superposition of several modes of energies $E({\bf{q}})$, $E({\bf{q}} \pm {\bf{k_{1}}})$, $E({\bf{q}} \pm {\bf{k_{2}}})$ and $E({\bf{q}} \pm {\bf{k_{3}}})$, with $E({\bf{k_{1}}}) = E({\bf{k_{2}}}) = E({\bf{k_{3}}})$. For a quasi-collinear arrangement with magnetic moments pointing mainly along the {\bf{b}}-axis, by assuming weakly $q_{a}$-dependent dynamical structure factors and a weak dispersion of excitations (as it is in BCAO, at least below 0.15 r.l.u.), one has $M_{{a^*}{a^*}} \approx M_{0}[N_{1}+(N_{2}+N_{3}){\cos}^{{2}}(2\pi/3)]\sin^2({\alpha})$ and $M_{bb} \approx M_{0}(N_{2}+N_{3}){\sin}^{2}(2\pi/3)$, where $M_{0}$ is the dynamical structure factor at $q_{a} \approx 0$ and $\alpha=({\bf{Q}}, {\bf{a}}^*)$, as usually.  In these relations, $N_{i}$ ($i=1,3)$ are the various $k$-domain populations, and we have made use of the fact that the fluctuations are transverse. As anticipated in section~\ref{Results} (SNP on the spin-dynamics), the $k$-domain structure partly restores the isotropy of magnetic fluctuations, and the experimental ratio at $q_{a} \approx 0$, $\frac{M_{a^{*}a^{*}}}{M_{bb}} \approx 1$ can be accounted if  $N_{1} \approx 1/3$, and $N_{2}+N_{3} \approx 2/3$ ($N_{1}-\frac{N_{2} +N_{3}}{2} \approx 0$). At the scattering vector ${\bf{Q}}=(0.27, 0, 4.67)$, for which $\sin^2({\alpha}) \approx 0.91$, one has the ratio $\frac{M_{a^{*}a^{*}}}{M_{bb}} \approx 0.91$, which does not reproduce quantitatively the experimental ratio $\frac{M_{a^{*}a^{*}}}{M_{bb}} \approx 0.84$. The additional reduction factor, larger and larger as $q_{a} \rightarrow 0.5$ (we found $\frac{M_{a^{*}a^{*}}}{M_{bb}} \approx 0.57$ at ${\bf{Q}}=(0.5, 0, 4.67)$, for which $\sin^2({\alpha}) \approx 0.75$) likely originates from the $q_{a}$-dependences of structure factors associated with the various domains. Unfortunately, any more quantitative comparison would require to dispose of more comprehensive calculations of magnetic excitation spectra in BCAO (including the $k$-domain effects), a task which is clearly out the scope of this paper.\\
For the sake of completeness, we have also investigated the effects of $k$-domains on the off-diagonal $P_{yx}$ and $P_{yz}$ terms. Since there are a priori no correlations between the fluctuation components belonging to two different $k$-domains, $P_{yz}$ will be given by the following relation:
 \begin{equation}
  P_{yz} = \frac{N_{1}M_{yz_{1}}^{+}+N_{2}M_{yz_{2}}^{+}+N_{3}M_{yz_{3}}^{+}}{N_{1}(M_{yy_{1}}+M_{zz_{1}})+N_{2}(M_{yy_{2}}+M_{zz_{2}})+N_{3}(M_{yy_{3}}+M_{zz_{3}})} \nonumber 
\end{equation}
in which the $M_{ij_{n}} (i, j = y, z)$ are the various magnetic cross-sections associated with domain ${n} (n=1,3)$. With the same assumptions, one can easily derive the expression giving the $P_{yz}$ element in the case of a scattering vector ${\bf{Q}}$ almost parallel to ${\bf{a}}^*$ (case $\sin(\alpha) \approx 0$), in presence of $k$-domains. Assuming $n_{+} - n_{-}$ identical for the three $k$-domains, after some trivial algebra, one obtains:
\begin{equation}
   P_{yz} \approx -2(n_{+}-n_{-})(N_{1}+\frac{N_{2}+N_{3}}{2})\tan(\beta)P_{0} \nonumber 
\end{equation}
which depends directly on the various $k$-domain populations. Taking into account that $N_{1} \approx \frac{N_{2}+N_{3}}{2} \approx \frac{1}{3}$, the effects of $k$-domains conduce to a reduction of the $P_{yz}$ element by a factor of about 2/3. With the above parameters, one calculates $P_{yz} \approx 0.08$, a value which is again in good quantitative agreement with the experimental results.
For ${\bf{Q}}$ almost parallel to ${\bf{c}}^{*}$ (case $\cos(\alpha) \approx 0$), the $P_{yx}$ and $P_{yz}$ polarization-matrix elements are given by the following relations:
 \begin{eqnarray}
  P_{yx} &\approx& -2(p_{+}-p_{-})(n_{+}-n_{-})(N_{1}+\frac{N_{2}+N_{3}}{2})\sin(\beta)\frac{\delta S_{c}}{\delta S_{a^{*}}} \nonumber  \\
  P_{yz} &\approx& \frac{\sqrt{3}}{2}(N_{2}-N_{3})P_{0} \nonumber
\end{eqnarray}
As for the previous case, $P_{yx} \approx 0$, and the $P_{yz}$ term is directly related to the difference $N_{2}-N_{3}$, and vanishes if the $k$-domains are equally populated. In order to account for the experimental value of $P_{yz}$ for ${\bf{Q}}=(0, 0, 4.67)$ ($\lesssim 0.027$, see Table~\ref{Table-3}), one has to assume that in BCAO ${\mid}N_{2}-N_{3}{\mid} \lesssim 3 \%$. \

To summarize the discussion of SNP results on the spin dynamics in BCAO, within the simple picture of precessing magnetic moments, the finite inelastic $P_{yz}$ and $P_{zy}$ matrix elements appear to be just a consequence of the existence of a {\textit{trivial}} correlation coupling the anisotropic IP and OP components, through the precession of canted Co$^{2+}$ magnetic moments. In BCAO, this correlation (indeed inherent to any spin wave) is enhanced by the strong planar character. Indeed, this {\textit{forced}} correlation reveals no newer pieces of information than those already known from the ground-state structure (e.g., the canting of magnetic moments). As disappointing it may appear, our results lead us to conclude unambiguously that the gaped low-energy excitations which have been observed in BCAO close to $q=0$ are spin waves associated with pseudo-ferromagnetic, weakly-coupled chains, the existence of an incommensurate ground-state structure being marginal in the problem. Finally, in spite of our experimental efforts, the understanding of their dispersion relation, especially the gap-energy value,  still remain an open problem. \

\section{Acknowledgements}

We would like to thank F. Tasset, S. V. Maleyev, B. Toperveg, O. Cepas and T. Ziman for very helpful discussions. 
\

\begin{table}
\centering 
\caption{Polarization matrices for the pure nuclear Bragg peaks (0, 0, 3), (0, 0, 9) and (1, 0, 1).}
\label{Table-1}
\begin{tabular}{ccccc}
    \hline
    \hline
    $\textbf{Q}   $  &  $P_{\alpha \beta}$     &       ${x}$     &     $y$      &     $z$   \\
    \hline
                               &  $x$ & 0.881(7) & 0.025(4) & 0.008(4) \\
   (0, 0, 3) & $y$ & -0.005(4) & 0.884(7) & 0.022(4) \\
                              & $z$ & -0.009(4) & 0.005(4) & 0.884(7) \\
  \hline
                              &  $x$ & 0.838(4) & 0.027(2) & 0.021(2) \\
   (0, 0, 9) & $y$ &  0.013(2) &  0.834(4) & 0.027(2) \\
                              & $z$ &  0.011(2) & -0.016(2) & 0.836(4) \\
    \hline
                           &  $x$ & 0.887(8) & 0.032(4) & 0.009(4) \\
   (1, 0, 1) & $y$ & -0.011(4) & 0.888(8) & 0.021(4) \\
                              & $z$ & -0.014(4) & -0.007(4) & 0.890(8) \\
  \hline
  \hline
\end{tabular}
\end{table}

\begin{table}
\centering 
\caption{Polarization matrices for the magnetic satellites (0, 0, 0)$^{+}=(0.265, 0, -1.33)$, (0, 0, 6)$^{+}=(0.265, 0, 4.67)$, 
(0, 0, 9)$^{+}=(0.265, 0, 7.67)$ and (1, 0, 1)$^{-}=(0.730, 0, 2.33)$.}
\label{Table-2}
\begin{tabular}{ccccc}
    \hline
    \hline
    $\textbf{Q}  $  &  $P_{\alpha \beta}$     &       ${x}$     &     $y$      &     $z$   \\
    \hline
                              &  $x$ & -0.914(10) & -0.007(5) & -0.010(5) \\
   (0.265, 0, -1.33)   & $y$ & -0.004(5) & -0.912(10) & -0.033(5) \\  
                              & $z$ & -0.026(5) &  -0.032(5) & 0.917(10) \\
  \hline
                              &  $x$ & -0.918(4) & 0.022(12) & 0.011(12) \\
   (0.265, 0, 4.67)   & $y$ & 0.033(12) & -0.914(4) & -0.065(7) \\ 
                              & $z$ & -0.016(5) & -0.080(12) & 0.910(5) \\
  \hline
                              &  $x$ & -0.919(5) & 0.035(14) & -0.001(14) \\
   (0.265, 0, 7.67)   & $y$ & -0.026(14) & -0.911(4) & -0.063(4) \\ 
                              & $z$ & -0.017(7) & -0.051(14) & 0.922(5) \\
  \hline
                              &  $x$ & -0.907(19) & -0.033(11) & 0.05(11) \\
   (0.730, 0, 2.33)    & $y$ & -0.022(9) & -0.895(19) & -0.168(12) \\  
                              & $z$ & 0.021(10) & -0.164(12) & 0.864(19) \\
  \hline
  \hline
\end{tabular}
\end{table}

\begin{table}
\centering 
\caption{Polarization matrices for several inelastic magnetic contributions.}
\label{Table-3}
\begin{tabular}{ccccc}
    \hline
    \hline
    $\textbf{Q}; {\Delta E} (meV) $  &  $P_{\alpha \beta}$     &       ${x}$     &     $y$      &     $z$   \\
    \hline
                              &  $x$ & -0.902(11) & 0.000(18) & -0.022(19) \\
   (0, 0, 4.67); 1.5 & $y$ & 0.012(19) & 0.022(13) & 0.027(18) \\   
                              & $z$ & -0.015(18) & 0.028(18) & 0.035(13) \\
  \hline
                              &  $x$ & -0.932(13) & 0.019(22) & 0.015(20) \\
   (0.27, 0, 4.67); 2.1 & $y$ & 0.008(22) & -0.154(15) & -0.007(13) \\  
                              & $z$ & 0.008(21) & 0.000(22) & 0.159(15) \\
    \hline
                              &  $x$ & -0.610(32) & * & -0.039(13) \\
   (0.73, 0, 0.8); 2.3 & $y$ & 0.015(13) & -0.620(30) & -0.074(20) \\
                              & $z$ & -0.052(22) & -0.074(20) & 0.745(33) \\
  \hline
  \hline
\end{tabular}
\end{table}

\begin{table}
\centering 
\caption{Off-diagonal matrix elements $Pxz$, $Pyx$, $Pyz$ and $Pzy$ at the scattering vectors {\bf Q}=(0.73, 0, 0.8) and {\bf Q}=(-0.73, 0, -0.8), for an energy transfer of 2.3 meV. $P^{a}_{\alpha \beta}$ and  $P^{s}_{\alpha \beta}$ are related to the antisymmetric and symmetric components, as explained in the text.}
\label{Table-4}
\begin{tabular}{ccccc}
    \hline
    \hline
    ${\alpha \beta}$  &   $P_{\alpha \beta}({\bf Q})$   &   $P_{\alpha \beta}(-{\bf Q})$  &  $P^{a}_{\alpha \beta}({\bf Q})$  &   $P^{s}_{\alpha \beta}({\bf Q})$   \\
    \hline
           $xz$           &  -0.039(13) & -0.031(12) & -0.004(8) & -0.035(8) \\
           $yx$           &   0.015(13) & 0.029(12) & -0.007(8) & 0.022(8) \\
           $yz$           & -0.035(18) & 0.088(17) & -0.062(13) & 0.026(13) \\
           $zy$           &  -0.047(18) & 0.129(18) & -0.088(14) & 0.042(14) \\
  \hline
  \hline
\end{tabular}
\end{table}

\section{References}

\begin{thebibliography}{00}

\bibitem{1D-HAF}
T. Yamada,
Prog. Theor. Phys. Jpn. {\textbf{41}}, 880 (1969). \\
DOI: https://doi.org/10.1143/PTP.41.880

\bibitem{1D-Haldane}
F.D.M. Haldane, 
Phys.Rev. Lett. {\textbf{50}}, 1153 (1983). \\
DOI: https://doi.org/10.1103/PhysRevLett.50.1153

\bibitem{HaldaneGap}
M.P. Nightingale and H.W.J. Bl{\"o}te, 
Phys. Rev. B {\textbf{33}}, 659 (1986). \\
DOI: https://doi.org/10.1103/PhysRevB.33.659

\bibitem{Haldane-J1J2}
F.D.M. Haldane,
Phys. Rev. B{textbf{25}}, 4925(R) (1982). \\
DOI: https://doi.org/10.1103/PhysRevB.25.4925

\bibitem{2legSpinLadder}
E. Dagotto and T.M. Rice,
Science {\textbf{271}}, 618 (1996). \\
DOI: https://doi.org/10.1126/science.271.5249.618

\bibitem{KT-Transition}
J.M. Kosterlitz and D.J. Thouless,
J. Phys. C: Solid State Phys. {\textbf(6)}, 1181 (1973). \\
DOI: https://doi.org/10.1088/0022-3719/6/7/010

\bibitem{Jose-Kadanoff}
J.V. Jos{\'{e}}, L.P. Kadanoff, S. Kirkpatrick and D.R. Nelson,
Phys. Rev. {\textbf{16}}, 1217 (1977). \\
DOI: https://doi.org/10.1103/PhysRevB.16.1217

\bibitem{graphene} 
D.S.L. Abergel, V. Apalkov, J. Berashevich, K. Ziegler, and T. Chakraborty,
Adv. Phys. {\textbf{59}}, 261 (2010) and references therein.\\
DOI: https://doi.org/10.1080/00018732.2010.487978

\bibitem{Iridate}
K.A Modic, T.E. Smidt, I. Kimchi, N.P. Breznay, A. Biffin, S. Choi, R.D. Johnson, R. Coldea, P. Watkins-Curry, G.T. McCandless, J.Y. Chan, F. Gandara, Z. Islam, A. Vishwanath, A. Shekhter, R.D. McDonald and J.G. Analytis,
Nature Communications {\textbf{5}}, 4203 (2014). \\
DOI: https://doi.org/10.1038/ncomms5203

\bibitem{alphaRuCl3}
A. Banerjee, J. Yan, J. Knolle, C.A. Bridges, M.B. Stone, M.D. Lumsden, D.G. Mandrus, D.A. Tennant, R. Moessner and S.E. Nagler
Science {\textbf{356}}, 1055 (2017). \\
DOI: https://doi.org/10.1126/science.aah6015

\bibitem{BCAO-deJongh} 
L.P. Regnault and J. Rossat-Mignod, 
 in \textit{Magnetic Properties of Layered Transition Metal Compounds}, 
 edited by L.J. de Jongh, (Kluwer Academic Press, 1990), p. 271-321.\\
 DOI: https://doi.org/0.1007/978-94-009-1860-3

\bibitem{Eymond-Durif69} 
S. Eymond, A. Durif and C. Martin,
C.R. Acad. Sci., {\textbf{286c}}, 1694 (1969). \\
http://gallica.bnf.fr/ark:/12148/bpt6k4802761/f1717.image

\bibitem{EymondDurif69} 
S. Eymond, C. Martin and A. Durif, 
Mat. Res. Bull. {\textbf{4}}, 595 (1969). \\
DOI: https://doi.org/10.1016/0025-5408(69)90120-21

\bibitem{BCAO-LPR83} 
L.P. Regnault, J. Rossat-Mignod, and J.Y. Henry,
J.Phys. Soc. Jpn. {\textbf{52}} Suppl., 1 (1983).

\bibitem{Dordevic-structure} 
T. Dordevic, 
Acta Cryst. {\textbf{E64}}, i58 (2008). \\
DOI: https://doi.org/10.1107/S1600536808025865

\bibitem{BCAO-suscep} 
L.P. Regnault, P. Burlet, and J. Rossat-Mignod, 
Physica B+C {\textbf{86-88}}, 660 (1977).\\
DOI: https://doi.org/10.1016/0378-4363(77)90635-0

\bibitem{BCAO-Excitations}
L.P. Regnault, J. Rossat-Mignod, J.Y. Henry, R. Pynn, and D. Petitgrand,
in \textit{Magnetic Excitations and Fluctuations},
edited by S.W. Lovesey, U. Balucani, F. Borsa, and V. Tognetti (Springer-Verlag, 1984), p.201-206  \\
DOI: https://doi.org/10.1007/978-3-642-82369-5

\bibitem{BCAO-chasp} 
L.P. Regnault, J. Rossat-Mignod, J. Villain, and A. de Combarieu, 
J. de Phys. {\textbf{C6}}, 759 (1978).\\
DOI: https://doi.org/10.1051/jphyscol:19786339

\bibitem{BCAO-HT} 
L.P. Regnault and J. Rossat-Mignod, 
J.Magn.Magn.Mat. {\textbf{14}}, 194 (1979).\\
DOI: https://doi.org/10.1016/0304-8853(79)90117-3

\bibitem{BCAO-energy(T)} 
L.P. Regnault, J.P. Boucher, J. Rossat-Mignod, J. Bouillot, R. Pynn, J.Y. Henry, and J.P. Renard, 
Physica B+C {\textbf{136}}, 329 (1986).\\
DOI: https://doi.org/10.1016/S0378-4363(86)80085-7

\bibitem{SNP-GEN} 
L.P. Regnault, H.M. R{\o}nnow, C. Boullier, J.E. Lorenzo, and C. Marin, 
Physica B {\textbf{345}}, 111 (2004).\\
DOI: https://doi.org/10.1016/j.physb.2003.11.035

\bibitem{LPA-textbook} 
L. Squires,
\textit{Introduction to the Theory of Thermal Neutron Scattering (3rd edition)} (Cambridge University Press, Cambridge, 2012) and references therein. \\
DOI: https://doi.org/10.1017/CBO9781139107808

\bibitem{Lovesey87}
S.W. Lovesey, 
\textit{Theory of Neutron Scattering from Condensed Matter}, vols. 1 and 2 (Clarendon Press, Oxford, 1984).

\bibitem{Balcar-Lovesey89}
E. Balcar and S.W. Lovesey, 
\textit{Theory of Magnetic Neutron and Photon Scattering} (Clarendon Press, Oxford, 1989).

\bibitem{Tasset-SNP-89}
F. Tasset, 
Physica B {\textbf{156-157}}, 627 (1989).\\
DOI: https://doi.org/10.1016/0921-4526(89)90749-7

\bibitem{LPR-SNP-04}
L.P. Regnault, B. Geffray, P. Fouilloux, B. Longuet, F. Mantegazza, F. Tasset, E. Leli\`evre-Berna, S. Pujol, E. Bourgeat-Lami, N. Kernavanois, M. Thomas and Y. Gibert, 
Physica B {\textbf{350}}, e811 (2004).\\
DOI: https://doi.org/10.1016/j.physb.2004.03.211

\bibitem{ELB-SNP-05}
E. Leli\`evre-Berna,  E. Bourgeat-Lami, P. Fouilloux, B. Geffray, Y. Gibert, K. Kakurai, N. Kernavanois, B. Longuet, F. Mantegazza, N. Nakamura, S. Pujol, L.P. Regnault, F. Tasset, M. Takeda, M. Thomas, and X. Tonon, 
Physica B {\textbf{356}}, 131 (2005).\\
DOI: https://doi.org/10.1016/j.physb.2004.10.063

\bibitem{JAEA-SNP-05}
M. Takeda , M. Nakamura, K. Kakurai, E. Leli\`evre-Berna, F. Tasset, and L.P. Regnault, 
Physica B {\textbf{356}}, 136 (2005).\\
DOI:  https://doi.org/10.1016/j.physb.2004.10.064

\bibitem{ELB-SNP-07}
E. Leli\`evre-Berna, P.J. Brown, F. Tasset, K. Kakurai, M. Takeda, and L.P. Regnault, 
Physica B {\textbf{397}}, 120 (2007).\\
DOI: https://doi.org/10.1016/j.physb.2007.02.076

\bibitem{Maleyev-SNP-63} S.V. Maleyev, V.G. Baryakhtar, and A. Suris, 
Sov. Phys. Solid State {\textbf{4}}, 2533 (1963).

\bibitem{Blume-SNP-63} 
M. Blume, 
Phys. Rev. {\textbf{130}}, 1670 (1963).\\
DOI: https://doi.org/10.1103/PhysRev.130.1670

\bibitem{JB-SNP-Chatterji}
P.J. Brown,
in \textit{Neutron Scattering from Magnetic Materials},
edited by T. Chatterji (Elsevier, 2006), p. 215-244 \\
DOI: https://doi.org/10.1016/B978-044451050-1/50006-9

\bibitem{LPR-SNP-Chatterji}
L.P. Regnault,
in \textit{Neutron Scattering from Magnetic Materials}
edited by T. Chatterji (Elsevier, 2006), p. 363-395 \\
DOI: https://doi.org/10.1016/B978-044451050-1/50009-4

\bibitem{Maleyev-Chiral}
S. Maleyev,
Physics Uspekhi, {\textbf{4}}, 569 (2002).

\bibitem{Trammell53}
G.T. Trammell, 
Phys. Rev. {\textbf{92}}, 1387 (1953). \\
DOI: https://doi.org/10.1103/PhysRev.92.1387

\bibitem{Selke-1980}
W. Selke and M.E. Fisher, 
Z. Physik B - Condensed Matter {\textbf{40}}, 71 (1980). \\
DOI: https://doi.org/10.1007/BF01295073

\bibitem{ANNNI-model}
J. Villain and P. Bak, 
J. Phys. (Paris) {\textbf{42}}, 657 (1981). \\
DOI: https://doi.org/10.1051/jphys:01981004205065700

\bibitem{Shirakura-2014}
T. Shirakura, F. Matsubara, and N. Suzukil, 
Phys. Rev. B {\textbf{90}}, 144410 (2014). \\
DOI: https://doi.org/10.1103/PhysRevB.90.144410

\bibitem{Rastelli79}
E. Rastelli, A. Tassi, and L. Reatto,
Physica B 97, 1 (1979). \\
DOI: https://doi.org/10.1016/0378-4363(79)90002-0

\bibitem{kagome-specifiheat}
P. Sindzingre, G. Misguich, C. Lhuillier, P. Bernu, L. Pierre, C. Waldtmann, and H.-U. Everts,
Phys. Rev. Lett. {\textbf{84}}, 2953 (2000). \\
DOI: https://doi.org/10.1103/PhysRevLett.84.2953

\bibitem{J1J2J3-Jussieu}
J.B. Fouet, P. Sindzingre, and C. Lhuillier, 
Eur. Phys. J. B {\textbf{20}}, 241(2001). \\
DOI: https://doi.org/10.1007/s100510170273

\bibitem{Takano06}
K. Takano,
Phys. Rev. B {\textbf{74}}, 140402(R) (2006). \\
DOI: https://doi.org/10.1103/PhysRevB.74.140402

\bibitem{cobalt2plus}
M.E. Lines,
Phys. Rev. {\textbf{131}}, 546 (1963). \\
DOI: https://doi.org/10.1103/PhysRev.131.546

 \bibitem{IC-Excitations}
T. Ziman and P.-A. Lindg{\aa}rd, 
Phy. Rev. B {\textbf{33}}, 1976 (1986). \\
DOI: https://doi.org/10.1103/PhysRevB.33.1976

\bibitem{Bishop-XY}
R.F. Bishop, P.H.Y. Li, and C.E. Campbell, 
Phys. Rev. B {\textbf{89}}, 214413 (2014). \\
DOI: https://doi.org/10.1103/PhysRevB.89.214413

\bibitem{Bishop-XXZ}
P.H.Y. Li, R.F. Bishop, and C.E. Campbell, 
Phys. Rev. B {\textbf{89}}, 220408(R) (2014). \\
DOI: https://doi.org/10.1103/PhysRevB.89.220408

\bibitem{Bishop-3}
R.F. Bishop, P.H.Y. Li, O. Gotze, J. Richter and C.E. Campbell, 
Phys. Rev. B {\textbf{92}}, 224434 (2015). \\
DOI: https://doi.org/10.1103/PhysRevB.92.224434

\bibitem{Bishop-4}
P.H.Y. Li and R.F. Bishop, 
Phys. Rev. B {\textbf{93}}, 214438 (2016). \\
DOI: https://doi.org/10.1103/PhysRevB.93.214438

\end{thebibliography}
\end{document}